\documentclass[twocolumn]{aastex631}

\usepackage{xcolor}
\usepackage{booktabs}
\usepackage{cprotect}
\usepackage{rotating}
\usepackage{orcidlink}
\usepackage{xspace}
\usepackage{multirow} 
\usepackage{enumitem}   
\usepackage{chemformula}
\usepackage[version=4]{mhchem} 
\usepackage{bm}
\usepackage{comment}
\usepackage{threeparttable}
\usepackage{tabularx}
\usepackage{hyperref}
\usepackage{footmisc}

\usepackage{graphicx}

\newcommand{\planet}{TOI-260\,b\xspace}

\newcommand{\tiberius}[1]{\texttt{Tiberius}\xspace}
\newcommand{\eureka}[1]{\texttt{Eureka!}\xspace}

\defcitealias{teskeMagellanTESSSurveySurvey2021}{MTS21}
\defcitealias{aldersonJWSTCOMPASSNIRSpec2024}{Alderson \& COMPASS et al. (2024)}
\defcitealias{wallackJWSTCOMPASSNIRSpec2024}{Wallack \& COMPASS et al. (2024)}
\defcitealias{alamJWSTCOMPASSFirst2025}{Alam \& COMPASS et al. (2025)}
\defcitealias{scarsdaleJWSTCOMPASS352024}{Scarsdale \& COMPASS et al. (2024)}

\begin{document}

\title{JWST COMPASS: A NIRSpec G395H Transmission Spectrum of Radius Valley-Dweller TOI-260\,b}

\author[0000-0002-7500-7173]{Annabella Meech}
\affiliation{Center for Astrophysics ${\rm \mid}$ Harvard {\rm \&} Smithsonian, 60 Garden St, Cambridge, MA 02138, USA}
\email{annabella.meech@cfa.harvard.edu}

\author[0000-0002-8518-9601]{Peter Gao}
\affiliation{Earth and Planets Laboratory, Carnegie Institution for Science, 5241 Broad Branch Road, NW, Washington, DC 20015, USA}

\author[0000-0003-0354-0187]{Nicole L. Wallack}
\affiliation{Earth and Planets Laboratory, Carnegie Institution for Science, 5241 Broad Branch Road, NW, Washington, DC 20015, USA}

\author[0000-0003-3204-8183]{Mercedes López-Morales}
\affiliation{Space Telescope Science Institute, 3700 San Martin Drive, Baltimore MD 21218, USA}

\author[0000-0002-2702-7700]{Dominic Oddo}
\affiliation{Department of Physics and Astronomy, University of New Mexico, 210 Yale Blvd NE, Albuquerque, NM, 87106, USA}

\author[0009-0008-2801-5040]{Johanna Teske}
\affiliation{Earth and Planets Laboratory, Carnegie Institution for Science, 5241 Broad Branch Road, NW, Washington, DC 20015, USA} \affiliation{The Observatories of the Carnegie Institution for Science, 813 Santa Barbara St., Pasadena, CA 91101, USA}

\author[0000-0003-2313-467X]{Diana Dragomir}
\affiliation{Department of Physics and Astronomy, University of New Mexico, 210 Yale Blvd NE, Albuquerque, NM, 87106, USA}


\author[0000-0003-2862-6278]{Angie Wolfgang}
\affiliation{Eureka Scientific, Inc.}

\author[0000-0002-0413-3308]{Nicholas Wogan}
\affiliation{NASA Ames Research Center, Moffett Field, CA 94035, USA}

\author[0000-0003-4328-3867]{Hannah R. Wakeford}
\affiliation{School of Physics, University of Bristol, HH Wills Physics Laboratory, Tyndall Avenue, Bristol BS8 1TL, UK}

\author[0000-0002-6721-3284]{Sarah E. Moran}
\affiliation{Space Telescope Science Institute, 3700 San Martin Drive, Baltimore, MD 21218}

\author[0000-0002-4207-6615]{James Kirk}
\affiliation{Department of Physics, Imperial College London, Prince Consort Road, London SW7 2AZ, UK}

\author[0000-0001-5253-1987]{Tyler A. Gordon}
\affiliation{Department of Astronomy and Astrophysics, University of California, Santa Cruz, CA 95064, USA}

\author[0009-0003-2576-9422]{Anna Gagnebin}
\affiliation{Department of Astronomy and Astrophysics, University of California, Santa Cruz, CA 95064, USA}

\author[0000-0003-1240-6844]{Natasha E. Batalha}
\affiliation{NASA Ames Research Center, Moffett Field, CA 94035, USA}

\author[0000-0002-7030-9519]{Natalie M. Batalha}
\affiliation{Department of Astronomy and Astrophysics, University of California, Santa Cruz, CA 95064, USA}

\author[0000-0001-8703-7751]{Lili Alderson}
\affiliation{Department of Astronomy, Cornell University, 122 Sciences Drive, Ithaca, NY 14853, USA}

\author[0000-0003-4157-832X]{Munazza K. Alam}
\affiliation{Space Telescope Science Institute, 3700 San Martin Drive, Baltimore, MD 21218, USA}

\author[0000-0002-8949-5956]{Artyom Aguichine}
\affiliation{Department of Astronomy and Astrophysics, University of California, Santa Cruz, CA 95064, USA}

\begin{abstract}

We present a JWST/NIRSpec G395H transmission spectrum of \planet, a $T_\mathrm{eq}\sim 490$\,K, $R_\mathrm{p} = 1.76\,R_\oplus$ planet.
The transmission spectrum is derived by combining two transit observations, collected as part of the JWST COMPASS program.
We achieved the same median transit depth precision of 37\,ppm in both visits, and a median precision of 26\,ppm when combining the spectroscopic light curves from the two visits.
Implementing a 30-pixel-wide ($R\sim 200$) spectroscopic binning scheme, we find that the transmission spectrum is mostly featureless, with a possible feature around 3.17\,$\mu$m.
We assess the significance of any features in the transmission spectrum with a suite of non-parametric models, which confirm the presence of a potential feature in the NRS1 bandpass and an offset between the NRS1 and NRS2 detectors.
To investigate the atmospheric composition of \planet, we run a series of PLATON retrievals.
We do not detect any clear molecular signatures, but the combined data from the two visits are sufficient to constrain the atmospheric metallicity to greater than $200\times$ solar, assuming no opaque deck $\lesssim2.5$\,mbar.
We also investigate causes of the potential feature near 3.17\,$\mu$m; while we find some compatible gaseous species and cannot fully discard an astrophysical origin, we suspect a systematics origin due to the variance in strength and position of the feature.
Overall, this look at \planet adds to the small sample of radius-valley planets, which already seem to show a diversity in their atmospheric compositions. Determining the true nature of these enigmatic planets will require a larger telescope time investment.

\end{abstract}

\keywords{Exoplanet atmospheric composition (2021); Exoplanet atmospheres (487); Exoplanets (498); Infrared spectroscopy (2285)}

\section{Introduction} 
\label{SECTION-1:INTRODUCTION}

The characterisation of small planets ($1-3R_\oplus$) has been a prominent aim of the exoplanet community in recent years due to their high occurrence rate relative to other sizes of planets, and the intriguing bimodality observed in their period-radius distribution as revealed by \textit{Kepler} \citep[][]{owenKeplerPlanetsTale2013,fultonCaliforniaKeplerSurvey2017,zengPlanetSizeDistribution2017}.
One of the main mechanisms proposed for driving this bimodality is atmospheric mass loss, driven by stellar XUV radiation (photoevaporation) or planet internal heat \citep[core-powered mass loss;][]{owenATMOSPHERESLOWMASSPLANETS2016,guptaSculptingValleyRadius2019,ginzburgCorepoweredMasslossRadius2018}. 
Models incorporating these mechanisms can match the observed radius-period distribution of small planets under the assumption that these planets are $\sim$Earth-like composition ``cores" that either fully lost or retained a fraction of an initially-accreted (``primordial") envelope of H/He gas. 
However, an alternate hypothesis that can explain the (less well-characterized) radius-mass distribution of these planets invokes a large water mass fraction in the intermediate-sized planets \citep{zeng2019}, and was originally suggested as evidence of formation beyond the water ice line and inward migration \citep[][]{luqueDensityNotRadius2022,venturini2020,burn2024}.
Whether or not formation beyond the water ice line is necessary has recently been called into question, due to possible geochemical interactions between primordial atmospheres and molten interiors causing water to be produced endogenically \citep[e.g.,][]{luo2024,kite2021,Schlichting2022,werlen2025}.
Since bulk densities yield degenerate compositional solutions \citep[e.g.,][]{rogers2024}, atmospheric characterisation is essential to help distinguish between these scenarios.

To unveil the true nature of these small worlds, the JWST COMPASS (Compositions of Mini-Planet Atmospheres for Statistical Study) Program (JWST Cycle 1 Program ID \#2512, PIs: N. E. Batalha and J. Teske, \citealt{batalhaSeeingForestTrees2021}) set out to pursue a spectroscopic survey of 12 exoplanets\footnote{While 11 of the planets are scheduled to be observed under this program, observations of the 12th target, TOI-175.02, are being collected under GTO program ID \#1224 (PI S. Birkmann).}, with radii spanning 1--3\,$R_\oplus$ \citep[][]{alamJWSTCOMPASSFirst2025,aldersonJWSTCOMPASSNIRSpec2024,aldersonJWSTCOMPASSNIRSpec2025,redaiJWSTCOMPASSNIRSpec2025,scarsdaleJWSTCOMPASS352024,TeskeJWSTCOMPASS2025,wallackJWSTCOMPASSNIRSpec2024}.
The main goal of this program is to uniformly characterize the targeted small planets, to conduct a demographic study and constrain population-level statistics.
The sample of targets were chosen from the subset of $R_p \leq 3$~R$_{\oplus}$ planets observed as part of the Magellan-TESS Survey (MTS; \citealt{teskeMagellanTESSSurveySurvey2021}; henceforth, \citetalias{teskeMagellanTESSSurveySurvey2021}), which aimed to constrain the masses of 30 small, TESS-detected planets using Magellan II/Planet Finder Spectrograph (PFS) radial velocities.
The COMPASS targets span a range of sizes, temperatures and stellar types ($R_\mathrm{p} \sim 1.2-2.6\,R_\oplus$; $T_\mathrm{eq}\sim500-1000\,$K; $T_\mathrm{eff}\sim3500-5200$\,K).
Of the 12 planets, there are 4 pairs of planets in the same system.
For full details on sample selection, see \citet{batalhaImportanceSampleSelection2022}.

To date, the COMPASS Program has published results for seven planets, of which none of their 3--5 micron transmission spectra have shown discernible atmospheric features. 
Two of the planets -- GJ~357\,b \citep[][]{redaiJWSTCOMPASSNIRSpec2025} and L\,98-59\,c \citep[][]{scarsdaleJWSTCOMPASS352024} -- are squarely in the super-Earth radius regime.
It is thus not surprising that their spectra are featureless as they are most likely among the sample to simply be bare rocks. 
Similarly, the irradiation experienced by L\,169-8\,b (at $T_\mathrm{eq}\sim980$\,K and $P=1.4$\,days) suggests that it would have lost any primordial H/He atmosphere, consistent with the observed featureless spectrum \citep[][]{alamJWSTCOMPASSFirst2025}.
The spectra of planets in the COMPASS Program with perhaps the most ambiguous composition lie in the intermediate radius range of $\sim1.7-2$ $R_{\oplus}$ and are more challenging to interpret -- they could have moderate metallicity atmospheres, lower metallicity atmospheres with clouds/haze, or even no atmospheres, given the current mass measurement precision
\citep[][]{aldersonJWSTCOMPASSNIRSpec2024,aldersonJWSTCOMPASSNIRSpec2025,teskeJWSTCOMPASSNIRSpec2025}. 
Finally, the largest COMPASS planet, TOI-836\,c, must have some kind of atmosphere, but again the data in hand cannot distinguish between low metallicity with clouds/haze or higher ($\gtrsim$ 200 $\times$ solar) metallicity \citep[][]{wallackJWSTCOMPASSNIRSpec2024}.
Considering both the COMPASS sample as well as other small planet transmission spectra published in the literature thus far \citep[e.g., GJ~486\,b, GJ~1132\,b, TOI-1685\,b;][]{moranHighTideRiptide2023,mayDoubleTroubleTwo2023,luqueDarkBareRock2024}, there are only a handful of cases where atmospheric molecular features have been detected \citep[K2-18\,b, GJ~9827\,d, TOI-270\,d, TOI-421\,b;][]{madhusudhanCarbonbearingMoleculesPossible2023,Benneke2024arXiv,piaulet-ghorayebJWSTNIRISSReveals2024,davenport2025}, although several planets do show intriguing signs of atmospheric escape from He 10830\AA\ and Ly-$\alpha$ observations \citep[TOI-776\,b and TOI-776\,c, GJ~3090\,b, TOI-836\,c;][]{Loyd2025,Ahrer2025,Zhang2025He}. 

Thus, the quest to use atmospheric observations to distinguish between small planet formation scenarios remains an active one. 
As noted above, of particular interest are the planets in the ``radius valley'' \citep[a dearth of planets with sizes $1.75-2\,R_\oplus$;][]{vaneylenAsteroseismicViewRadius2018} -- neither bona fide super-Earths or sub-Neptunes. 
In this paper, we present the COMPASS observations of one of these planets.
\planet\ (HIP~1532\,b, TIC~37749396\,b) is a $R_\mathrm{p}=1.76\,R_\oplus$ planet\footnote{Considering the mass of TOI-260, \citet{venturiniFadingRadiusValley2024} would predict a valley minima at $R_\mathrm{p}=1.71\pm0.06\,R_\oplus$.} \citep[$P=13.478$\,days; Table~\ref{table:system_params};][]{teskeMagellanTESSSurveySurvey2021}, whose transit has been detected by the TESS SPOC pipeline in sectors 3, 42, and 70 \citep[][]{guerreroTESSObjectsInterest2021,stassunVizieROnlineData2019}.
It orbits a bright, early-type M dwarf \citep[Table~\ref{table:system_params};][]{stassunRevisedTESSInput2019,teskeMagellanTESSSurveySurvey2021}.
Several papers have measured the mass and/or radius of \planet, with results ranging from $M_\mathrm{p}=3.3-4.23\,M_\oplus$, and $R_\mathrm{p}=1.47-1.76\,R_\oplus$ \citep[][]{hobsonThreeSuperEarthsPossible2024,polanskiTESSKeckSurveyXX2024,teskeMagellanTESSSurveySurvey2021}.
In this work, we adopt the planet mass and radius published by \citetalias{teskeMagellanTESSSurveySurvey2021}, as a conservative estimate (see Table~\ref{table:system_params}).

Here, we present the first transmission spectrum of \planet.
Our JWST/NIRSpec G395H observations are described in \S\ref{SECTION-2:OBSERVATIONS}.
The data reduction and light curve detrending is detailed in \S\ref{SECTION-3:REDUCTION}.
Then, the near-infrared transmission spectrum of \planet\ is shown in \S\ref{SECTION-4:TRANMISSION_SPECTRUM}, and the interpretation and modelling is presented in \S\ref{SECTION-5:MODELLING}.
Finally, our results are discussed and summarised in \S\ref{SECTION-6:DISCUSSION} and \S\ref{SECTION-7:conclusions}.

\begin{table}
\begin{threeparttable}
   \caption{Adopted system parameters for the TOI-260 system.}
   \label{table:system_params}
    \begin{tabularx}{\columnwidth}{l c c}
    \toprule
    Parameter & Value & Reference \\ 
    \midrule
        \multicolumn{3}{c}{Star} \\
        \hline
        $M_*\,[M_\odot]$ & $0.630 \pm0.081$ & \citetalias{teskeMagellanTESSSurveySurvey2021} \\
        $R_*\,[R_\odot]$ & $0.618 \pm0.060$ & \citetalias{teskeMagellanTESSSurveySurvey2021}\\
        $T_\mathrm{eff}\,[\mathrm{K}]$ & $4026\pm14$ & \citet{hobsonThreeSuperEarthsPossible2024} \\
        $\log g \, [\mathrm{cgs}]$ & $4.45\pm0.05$ & \citet{hobsonThreeSuperEarthsPossible2024}\\
        metallicity & $-0.47\pm0.03$ & \citet{hobsonThreeSuperEarthsPossible2024}\\
        \hline
        \multicolumn{3}{c}{Planet b} \\
        \hline
        $P_\mathrm{orb}\,[\mathrm{days}]$ & $13.478048\pm0.005188$ & \citetalias{teskeMagellanTESSSurveySurvey2021} \\
        $M_\mathrm{p}\,[M_\oplus]$ & $3.97^{+2.39}_{-2.23}$ & \citetalias{teskeMagellanTESSSurveySurvey2021}\\
        $R_\mathrm{p}\,[R_\oplus]$ & $1.76\pm0.30$ & \citetalias{teskeMagellanTESSSurveySurvey2021} \\
        $T_\mathrm{eq}\,[\mathrm{K}]$ & $489\pm26$ & Derived using \citetalias{teskeMagellanTESSSurveySurvey2021} $\dagger$ \\
    \bottomrule
    \end{tabularx}
    \begin{tablenotes}
    \item[$\dagger$] Derived assuming an albedo of zero.
    \end{tablenotes}
    \end{threeparttable}
\end{table}  

\begin{figure*}
\centering
\includegraphics[width=\textwidth]{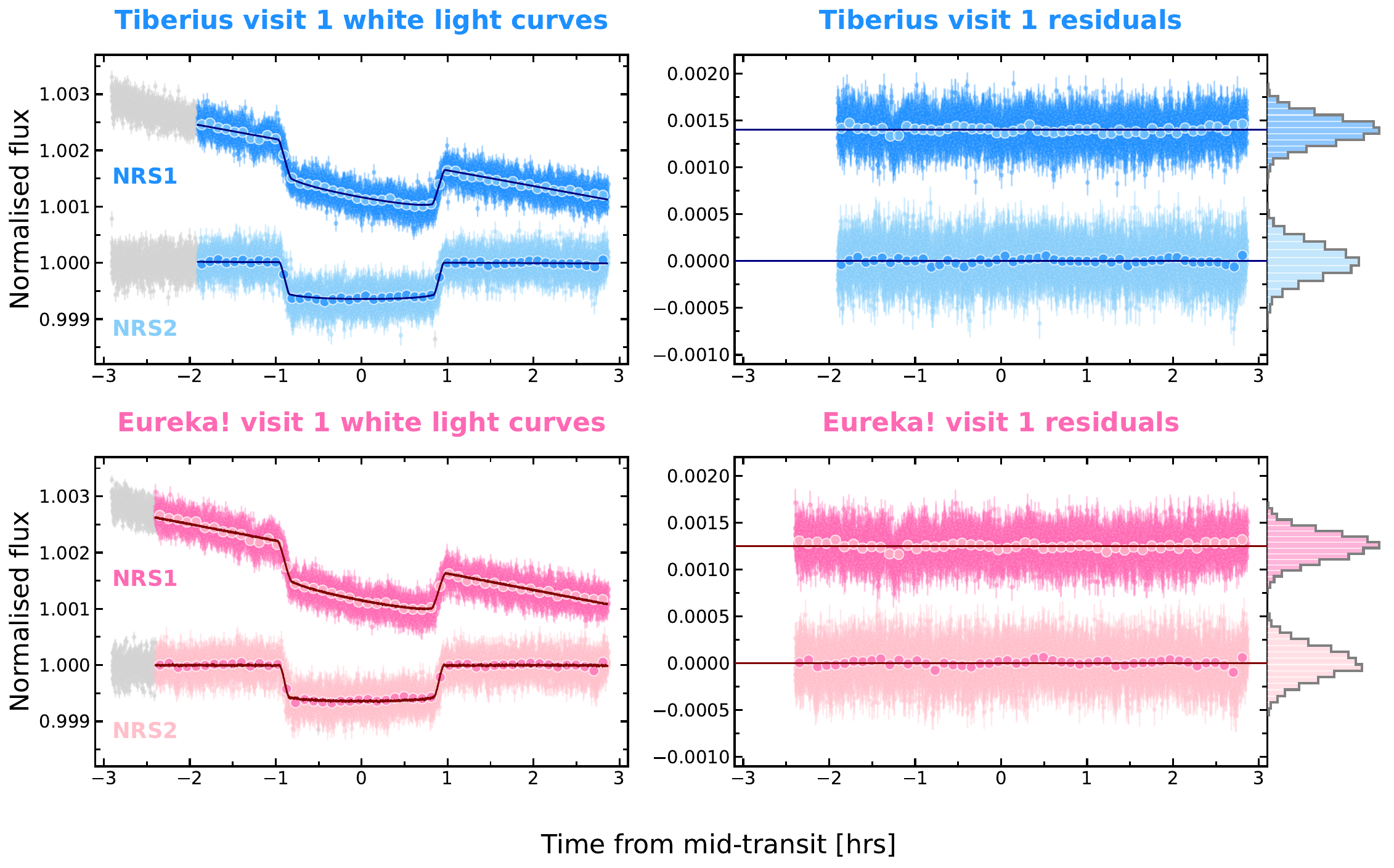}
\cprotect\caption{\textbf{Left:} The normalised \planet\ white light curves from visit 1 as extracted and detrended with \textbf{(top row)} \texttt{Tiberius} and \textbf{(bottom row)} \texttt{Eureka!}, centered on the fitted mid-transit time. 
The NRS1 light curve is shown in the darker shade (and is offset from unity for clarity) above the NRS2 light curve in the lighter shade. 
The best-fit systematics + transit models are overplotted.
The binned light curves are shown in alternate colors, and data points that were clipped prior to fitting are shown in grey, for reference only.
\textbf{Right:} The associated residuals for each visit and detector. We attach the histograms of the residuals to the right.}
\label{fig:WLC_v1}
\end{figure*}

\section{JWST/NIRSpec Observations} 
\label{SECTION-2:OBSERVATIONS}

Under the cycle~1 JWST program ID \#2512, we observed two transits of \planet\ with JWST/NIRSpec \citep{jakobsenNearInfraredSpectrographNIRSpec2022} on the 17th and 30th November 2023. 
The observations were scheduled based on transit ephemerides confirmed with CHEOPS and TESS data; we analyse these data in Appendix~\ref{APPENDIX-2:cheops}.
The JWST/NIRSpec observations were taken using the high-resolution G395H grating, with the SUB2048 subarray, F290LP filter, S1600A1 slit, and the NRSRAPID readout pattern.
The G395H grating affords spectra from 2.8--5.14\,microns across the detectors NRS1 and NRS2.
There exists a gap between these detectors, from 3.72--3.82\,microns. 
The observations were taken with the NIRSpec Bright Object Time Series (BOTS) mode. 
Each 5.8\,hr exposure consisted of 5718 integrations with 3 groups per integration.
These captured the entirety of the transit as well as sufficient out-of-transit baseline on both sides of the transit, to enable accurate detrending.
The data were delivered in four segments to the MAST database, and are publicly available at \url{https://mast.stsci.edu/}.

\section{Data Reduction}
\label{SECTION-3:REDUCTION}

We performed two independent reductions of the data using the open-source pipelines \tiberius\ \footnote{\url{https://tiberius.readthedocs.io/en/latest/}} \citep{kirkRayleighScatteringTransmission2017,kirkACCESSLRGBEASTSPrecise2021} and \eureka\ \footnote{\url{https://eurekadocs.readthedocs.io/en/latest/}} \citep{bellEurekaEndEndPipeline2022}.
In both reductions, we processed the NRS1 and NRS2 data separately.
We detail each in turn below, and a summary is provided in Appendix~\ref{APPENDIX-1:reductions}, Table~\ref{table-app:reduction-details}.

\begin{figure*}
\includegraphics[width=\textwidth]{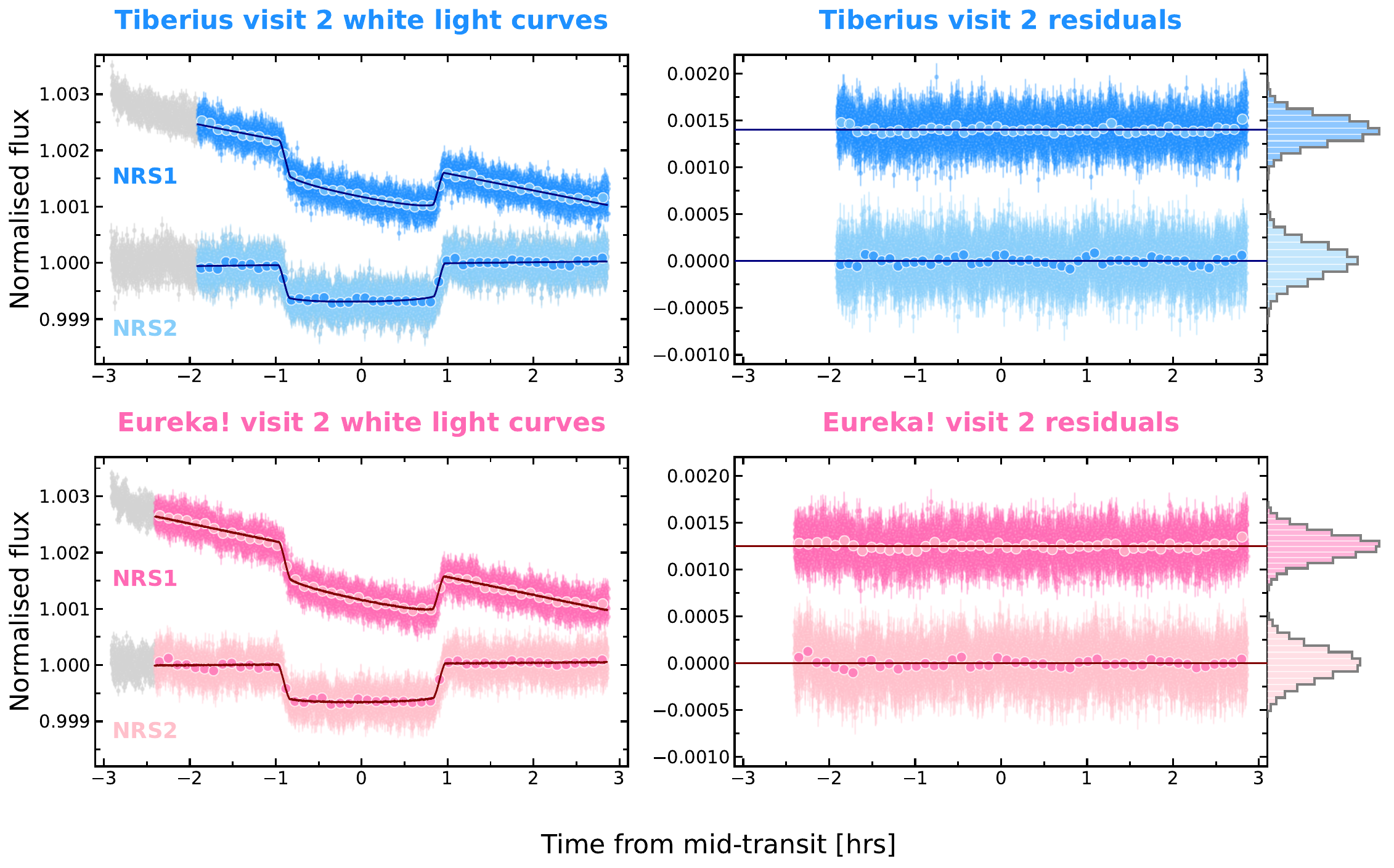}
\caption{The \planet\ white light curves from visit 2; same format as Fig.~\ref{fig:WLC_v1}.}
\label{fig:WLC_v2}
\end{figure*}
\begin{table*}[t]
\centering
    \caption{Retrieved planet parameters from the best fits of the white light curves. }
    \begin{threeparttable}
        
    \footnotesize
    \begin{tabular}{l c c c c c}
    
    \toprule
      & & $T_\mathrm{mid}$\,[BJD$_\mathrm{TDB}$] & $a/R_*$ & $i$\,[deg] & $R_\mathrm{p}/R_*$  \\
     
    \midrule
    Literature value & & $60264.941300\pm0.000858$\tnote{$\dagger$} & $33.92\pm 3.88$\tnote{$\ddagger$} & $88.75\pm 0.30$\tnote{$\ddagger$} & $0.0261\pm 0.0017$\tnote{$\ddagger$}\\
     & & $60278.417200\pm0.000862$\tnote{$\dagger$} \\
    \midrule
    \verb+Tiberius+ & Visit 1 NRS1 & $60264.939548{\substack{+0.000077\\-0.000079}}$& $33.07^{+1.04}_{-0.96}$& $88.58\pm 0.06$& $0.027706\pm0.000120$ \\ 
    & Visit 1 NRS2  & $60264.939544\pm0.000096$ & $36.21{\substack{+1.88\\-1.72}}$ & $88.77\pm0.10$ & $0.025472{\substack{+0.000153\\- 0.000154}}$ \\ 
    & Visit 2 NRS1 & $60278.415478{\substack{+0.000076\\-0.000077}}$ & $35.21\pm1.16$& $88.71\pm0.07$ & $  0.026788{\substack{+0.000124\\-0.000122}}$\\ 
    & Visit 2 NRS2  & $60278.415752{\substack{+0.000099\\-0.000100}}$ & $34.68{\substack{+1.69\\-1.47}}$ & $88.69{\substack{+0.10\\-0.09}}$ & $0.025859{\substack{+0.000146\\-0.000148}}$\\
    
    \midrule
    \verb+Eureka!+ & Visit 1 NRS1  & $60264.939446 \pm0.000080$ & $31.93\pm 0.97$  & $88.51 \pm 0.07$  & $0.027983 \pm 0.000115$ \\
    & Visit 1 NRS2  & $60264.939374 \pm 0.000095$  & $37.20 \pm 2.48$ & $88.83 \pm 0.13 $& $0.025124 \pm 0.000153$ \\
    & Visit 2 NRS1  & $60278.415421 \pm 0.000078$ & $34.63 \pm 1.25$ & $88.68 \pm 0.08$ & $0.026967 \pm 0.000122$ \\
    & Visit 2 NRS2 & $60278.415627 \pm 0.000097$ & $33.44 \pm1.37$  & $88.61 \pm 0.09$ & $0.026104 \pm 0.000125$ \\
    \bottomrule
    \end{tabular}
    \begin{tablenotes}
        \item[$\dagger$] Propagated for visit 1 and visit 2, using values from the CHEOPS + TESS ephemeris derived in Appendix~\ref{APPENDIX-2:cheops} (Table~\ref{table-app:cheops_params});
        \item[$\ddagger$] \cite{teskeMagellanTESSSurveySurvey2021}.
    \end{tablenotes}
    \end{threeparttable}
    \label{table:retrieved-params}

\end{table*}

\subsection{\texttt{Tiberius}}
\label{section-3:redution_tiberius}
We began the \tiberius\ reduction with the stage~1 processing of the uncalibrated detector images (\texttt{uncal.fits}), using the \texttt{jwst} pipeline\footnote{\url{https://jwst-pipeline.readthedocs.io/en/latest/}} \citep[v1.13.4;][]{bushouse_2025_16280965}.
We ran all of the default steps, including the jump step, with a rejection threshold of $10\sigma$, finding the noise level to be slightly lower than with this step excluded.
For the removal of $1/f$ noise, we performed a custom correction at the group level, as used in other \tiberius\ implementations \citep[see e.g.,][]{redaiJWSTCOMPASSNIRSpec2025}.
We defined the spectral trace and subtracted the median count level of the pixels outside the trace, in each column.
To finish the stage~1 processing, we implemented the standard ramp fit and gain scale corrections to combine the 3 groups per integration.

We proceeded to extract stellar spectra from the corrected NIRSpec detector images.
First, to correct for cosmic rays, we flag any given pixel that has a count level more than 4-sigma from the running median value along the time axis.
We interpolate over these, as well as any NaN pixels (which are a combination of the masked bad pixels and the \textsc{do\_not\_use} pixels), using the pixels to the left and right.
For each integration, the spectral extraction routine in \tiberius\ finds the center of the aperture trace by fitting each column with a Gaussian profile (along the cross-dispersion axis).
This trace profile is then smoothed by fitting the centroids of the Gaussians with an order-4 polynomial.
We summed the counts over a constant aperture width of 16 pixels (found to optimize signal-to-noise) centered on the fitted trace, to obtain a time-series of stellar spectra.
We extracted the stellar spectra over pixels [600--2040] inclusive for NRS1, and [5--2040] for NRS2, along the dispersion axis.

For the wavelength solution, we use that provided by the \texttt{assign\_wcs} step from stage~2 of the \texttt{jwst} pipeline.
To finish the processing of the stellar spectra, we resampled them using \texttt{pysynphot}, to ensure they were aligned onto a universal wavelength solution.
We cross-correlated each spectrum with the first spectrum in the time-series to deduce the optimal global shift in the dispersion direction\footnote{These shifts are saved as a \texttt{yposition} array for future detrending.}. 
The average shift was less than 7.5\% of a pixel width.

The white light curves (WLCs) are generated by integrating the counts across wavelength, at each timestamp.
Our \tiberius\ white light curves are shown in the top panels of Fig.~\ref{fig:WLC_v1} and Fig.~\ref{fig:WLC_v2}; we achieve median photometric precisions in the white light curves of 143\,ppm, 177\,ppm for NRS1 and NRS2 in visit 1, and 142\,ppm, 179\,ppm in visit 2.
We also construct spectroscopic light curves (SLCs) by summing the flux counts over pre-defined wavelength bins; we adopt the 30-pixel binning scheme used in previous COMPASS papers \citep[$\mathrm{R}\sim 200$; $\sim0.02\,\mu$m bin widths; 42 and 64 bins for NRS1 and NRS2 respectively;][]{aldersonJWSTCOMPASSNIRSpec2024,alamJWSTCOMPASSFirst2025,wallackJWSTCOMPASSNIRSpec2024,scarsdaleJWSTCOMPASS352024}.

We fit the light curves from each of the detectors and each of the visits separately, with a model that simultaneously accounts for both systematics and the transit light curve.
For the latter, we use the \texttt{batman} package\footnote{\url{http://lkreidberg.github.io/batman/docs/html/index.html}} \citep{kreidbergBatmanBAsicTransit2015} and a quadratic limb-darkening relation for all light curves.
We explored the parameter space with Markov Chain Monte Carlo (MCMC) sampling, using the \texttt{emcee}\footnote{\url{https://emcee.readthedocs.io/en/stable/}} package \citep[][]{DFM2013}.
For all light curves, we fixed the limb-darkening coefficients to values calculated with \texttt{ExoTiC-LD} \citep[][]{grantExoTiCLDThirtySeconds2024}.
For this, we chose to use the \textsc{stagger} grid of 3D stellar models \citep{magicStaggergridGrid3D2015}, retrieving the models corresponding to the stellar parameters given in Table~\ref{table:system_params}.
We also fixed the planet orbital period to $P=13.475832$\,days, and assume a circular orbit ($e=0$).
Having trialed different models for each detector and visit separately, we found that simple linear models were sufficient to detrend all light curves, with our chosen systematics model of the form:
\begin{equation}
    S(t) = p_1 + (p_2 \times T)
\label{eqn:systematics_model}
\end{equation}
where $T$ is the time array, and $p_N$ the polynomial coefficients which are left as free parameters. 
To remove the initial settling ramps, we exclude up to the first 60\,mins (992 integrations) for both visit 1 and visit 2  -- these integrations are greyed out in Fig.~\ref{fig:WLC_v1} and Fig.~\ref{fig:WLC_v2}.
This left us with 2.8\,hrs of out-of-transit baseline (1.5$\times$ the transit duration).

\begin{figure}[t]
    \includegraphics[width=0.45\textwidth]{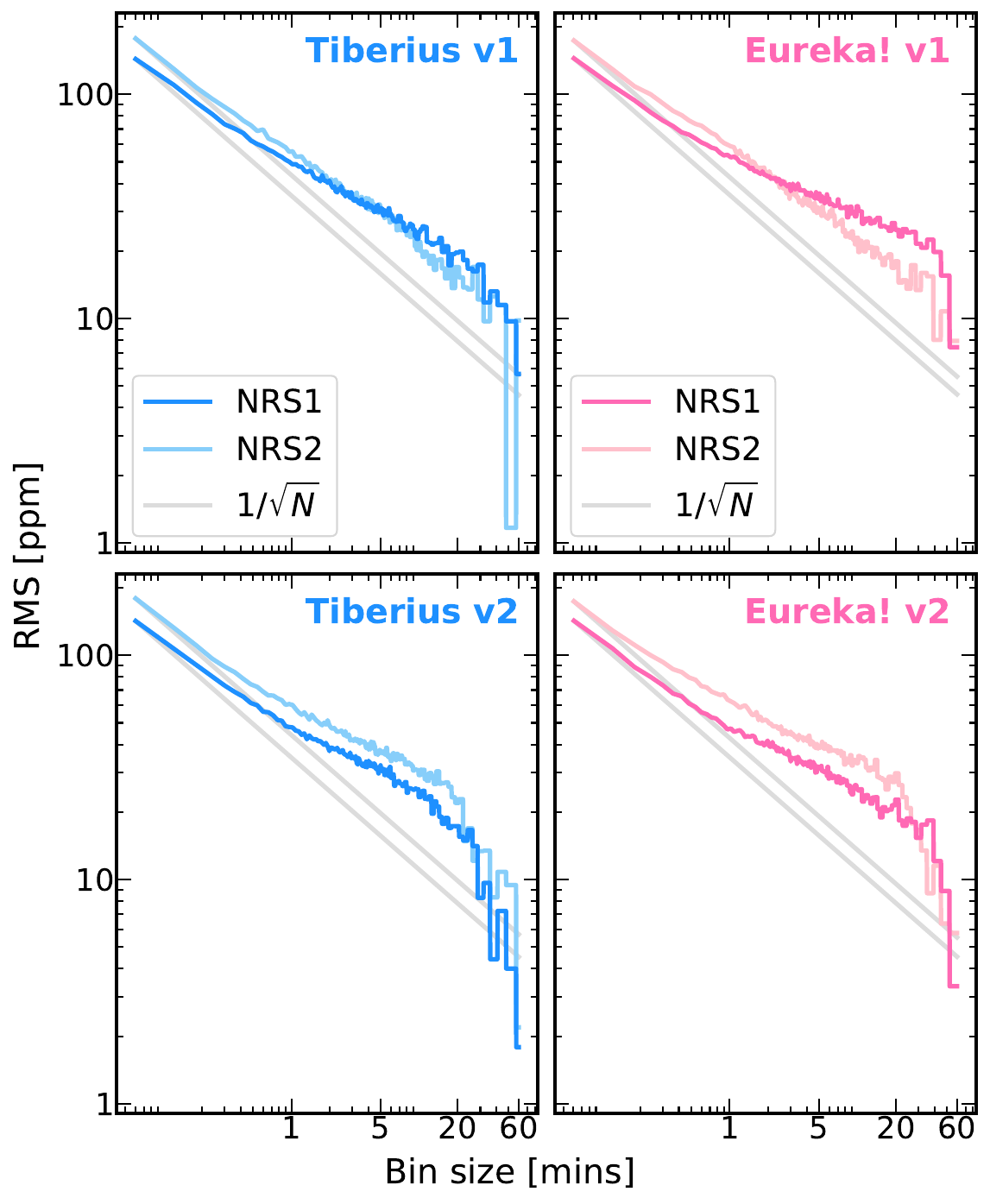}
    \caption{The RMS of the binned white light residuals for \textbf{(top row)} visit 1 and \textbf{(bottom row)} visit 2. \texttt{Tiberius} and \texttt{Eureka!} are shown on the left and right respectively.
    The expectation from pure Gaussian noise is indicated for reference in grey.}
    \label{fig:allan_WLC}
\end{figure}

For each WLC fitting, we had six free parameters: the mid-transit time, scaled semi-major axis $a/R_*$, inclination $i$, scaled planet radius $R_\mathrm{p}/R_*$ and the polynomial coefficients $p_N$.
We fixed the limb-darkening coefficients to $u_1=0.1186$, $u_2=0.1958$ for NRS1 and $u_1=0.098$, $u_2=0.1582$ for NRS2.
To explore the parameter space, we deployed 30 walkers per free parameter, and ran an initial burn-in of 10,000 steps.
Taking the resulting best-fit model, we rescale the photometric uncertainties to give $\chi_\nu^2=1$ and then run an additional 10,000 production steps.
The white light curves, their fitted models, and associated residuals are shown in Fig.~\ref{fig:WLC_v1} and Fig.~\ref{fig:WLC_v2}.
We show the RMS of the binned WLC residuals in Fig.~\ref{fig:allan_WLC}, compared to that expected from pure Gaussian noise, in the absence of other noise sources.
The posterior distributions of the fitted parameters are shown in Appendix~\ref{APPENDIX-1:reductions}, Fig.~\ref{fig-app:wlc-corner}.
The NRS1 light curves exhibit a particularly strong time-dependent slope, $49\times$ that measured for NRS2 in visit~1 and $-18\times$ the NRS2 slope in visit~2.
This is an unresolved, characteristic systematics effect seen in NIRSpec light curves \citep{espinozaSpectroscopicTimeseriesPerformance2022}.

We note that we tested a range of parameterized systematics models, including a combination of exponential and linear terms, and higher-order polynomials.
While the retrieved parameters (and therefore the absolute transit depths) varied slightly upon model selection, the resulting shape of the transmission spectrum (relative transit depths) remained consistent.
We opted for the linear model for consistency between detectors and broadband versus spectroscopic light curves.
We also tested the inclusion of additional metrics \texttt{xpositions}\footnote{The smoothed trace center pixel positions in the cross-dispersion direction, i.e., the position of the aperture trace, averaged over wavelength.}, \texttt{ypositions}\footnote{Global shift of the spectrum along the dispersion direction.} and \texttt{fwhm}\footnote{The full width half maximum of the Gaussians fitted along the cross-dispersion direction, i.e. the width of the aperture trace. We note that this was only an output of the trace-locating stage of \tiberius, and we used a constant-width aperture for the spectral extraction. For detrending of the WLCs, the 2D \texttt{fwhm} array is averaged over wavelength, unless otherwise specified.} in the systematics models, finding that these additional terms offered no improvement over the linear-in-time-only model.
Moreover, the retrieved parameters were similar to that of the adopted model.
We provide more detail regarding these tests, and the retrieved parameters from the different models trialed, in Appendix~\ref{APPENDIX-1:reductions}, Table~\ref{table-app:WLC_params}.

\begin{figure*}
\centering

\includegraphics[width=\textwidth]{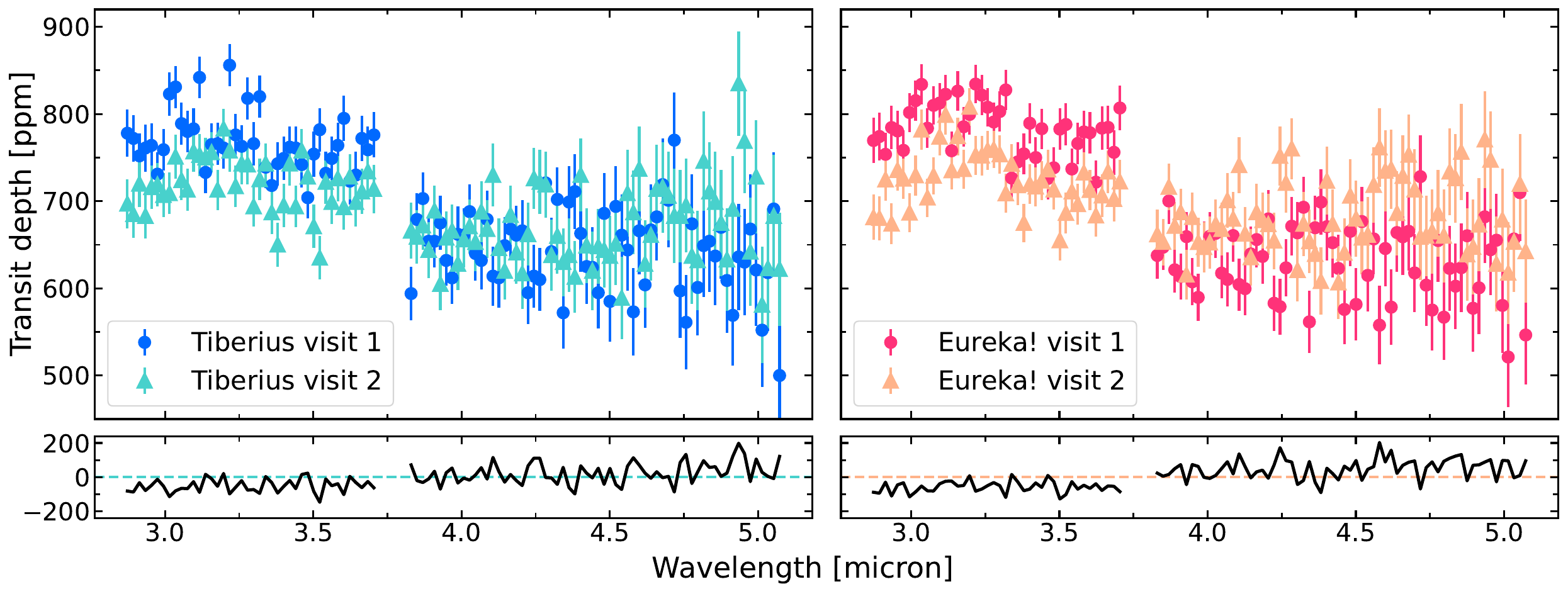}
\includegraphics[width=\textwidth]{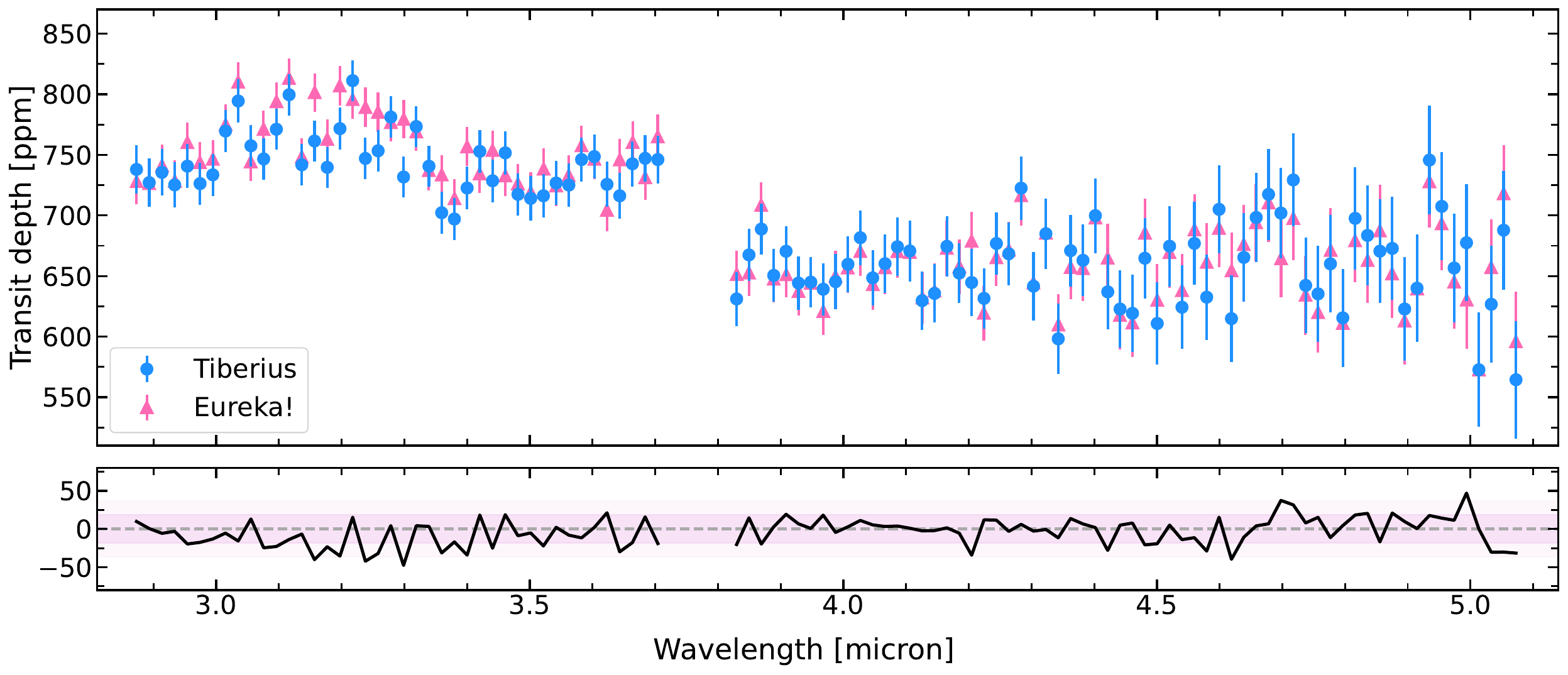}
\cprotect\caption{\textbf{Top row:} The transmission spectra of \planet, extracted with \textbf{(left)} \texttt{Tiberius} and \textbf{(right)} \texttt{Eureka!}. The spectra from the two visits are distinguished in different shades, and their difference plotted below in black (ppm).\\
\textbf{Bottom row:} The weighted average transmission spectra of \planet, from \texttt{Tiberius} (blue) and the transit depths from the jointly fit SLCs from \texttt{Eureka!} (pink).
The difference between the two joint spectra is plotted below in black. 
The median difference of 3.2\,ppm is marked with the grey dashed line, with $1\sigma$ and $2\sigma$ regions shaded.}

\label{fig:transmission_spectra}
\end{figure*}

The median retrieved system parameters from the individual WLC fits are given in Table~\ref{table:retrieved-params}.
These were then fixed when fitting the spectroscopic light curves (SLCs).
We fixed the limb-darkening coefficients to those computed for each spectroscopic bin, again using \texttt{ExoTiC-LD}.
For each visit, we used the same systematic + transit model across all spectroscopic bins (Table~\ref{table-app:reduction-details}).
For the fitting of each spectroscopic light curve, we used MCMC with three walkers per free parameter, and otherwise the same setup as the WLCs.
There were three free parameters per light curve: the scaled planet radius $R_\mathrm{p}/R_*$, and the linear polynomial coefficients, $p_N$.
The RMS of the binned residuals of these fits are shown in Appendix~\ref{APPENDIX-1:reductions}, Fig.~\ref{fig-app:allan_SLC}.
We construct the transmission spectrum by recovering the transit depth of each of the SLC transit models.
The resulting transmission spectra from each visit are presented in blue in Fig.~\ref{fig:transmission_spectra}.

\begin{figure}
    \centering
    \includegraphics[width=\columnwidth]{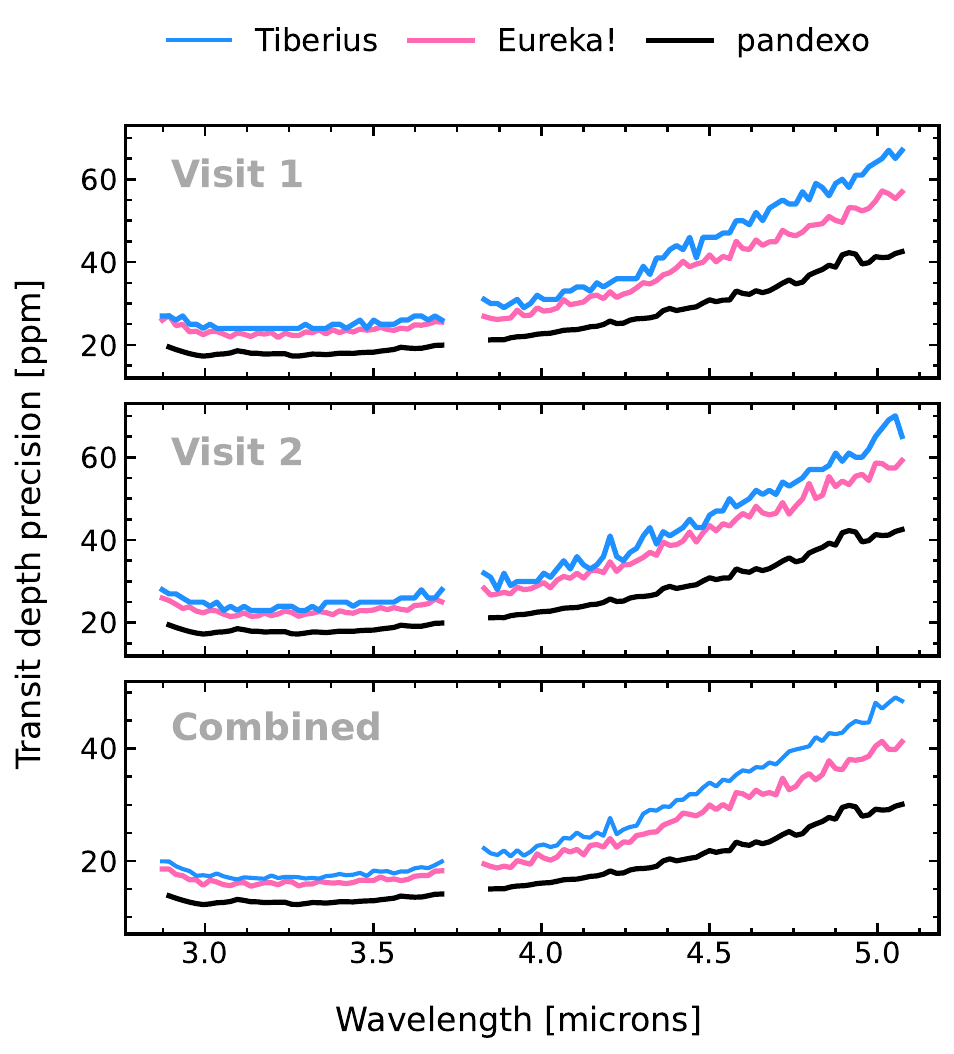}
    \cprotect\caption{The transit depth precisions measured by \verb|Tiberius| (blue) and \verb|Eureka!| (pink) compared to the predicted uncertainties from our \verb|pandexo| simulations (black).}
    \label{fig:spectral_precision}
\end{figure}

\subsection{\texttt{Eureka!}}
We independently employed the \verb+Eureka!+ pipeline \citep{bellEurekaEndEndPipeline2022} to reduce the NIRSpec time-series data. We utilize v0.10 of the \texttt{Eureka!} pipeline in a similar manner to other COMPASS papers \citep[e.g.][]{aldersonJWSTCOMPASSNIRSpec2024, alamJWSTCOMPASSFirst2025, scarsdaleJWSTCOMPASS352024, wallackJWSTCOMPASSNIRSpec2024}, and briefly summarize the process here. \texttt{Eureka!} is an end-to-end pipeline for the reduction of HST and JWST observations. When reducing JWST data, \texttt{Eureka!} acts as a wrapper around the \texttt{jwst} pipeline. We use the default parameters for the \texttt{jwst} pipeline, with the exception of a 15$\sigma$ jump detection threshold and the \texttt{Eureka!} group level background subtraction. We use version 1.11.4 of the \texttt{jwst} pipeline with context map jwst\_1293.pmap (see Appendix~\ref{APPENDIX-1:reductions}, Table~\ref{table-app:reduction-details} for a summary of reduction choices).

Stage 3 of \texttt{Eureka!} allows for different reduction parameters, so in order to determine the optimal combination of these reduction choices, we iterate over different values for the background and extraction apertures, the polynomial order for an additional background subtraction, and the sigma threshold for the outlier rejection for the optimal extraction, in a similar way as described in other COMPASS papers \citep[e.g.][]{wallackJWSTCOMPASSNIRSpec2024}. We consider extraction apertures of 4-8 pixel half-widths, background apertures of 8-11 pixels, sigma threshold for the outlier rejection during the optimal extraction of 10 and 60, and either an additional full frame or column-by-column background subtraction. We select the version of the reduction that minimizes the median absolute deviation. By this metric, we favor 4 pixel half-widths for the extraction and 8 pixel background apertures for both detectors for visit 2 and visit 1 NRS2, and a 5 pixel half-width and 9 pixel background aperture for visit 1 NRS1. 
We favored 10$\sigma$ outlier rejection thresholds for visit 1 NRS2 and both detectors for visit 2, and 60$\sigma$ for visit 1 NRS1. 
We favored an additional column-by-column background subtraction for all detectors and visits except for visit 2 NRS2.

We then extract 30-pixel wide spectroscopic light curves from 2.863--3.714\,$\mu$m  for NRS1 and 3.812--5.082\,$\mu$m for NRS2. 
To fit the light curves, we again use a custom fitting routine over the inbuilt \texttt{Eureka!} fitter to allow increased flexibility, but we nevertheless call this reduction the `\texttt{Eureka!} reduction' for simplicity, as we have done in previous COMPASS papers. 
For both the spectroscopic and white light curves, we first remove the first 30\,mins (500 integrations) from the time series to account for initial settling, and we iteratively trim 3$\sigma$ outliers from a 50-point rolling median three times and always treat the two detectors separately. 

For the white light curves, we fit for the $i$, $a/R_{\star}$, $T_{0}$, and $R_{p}/R_{\star}$ using \texttt{batman} \citep{kreidbergBatmanBAsicTransit2015}. Simultaneously, we fit a systematic noise model of the form 
\begin{equation}
S(t) = p_{1} + (p_{2}\times T) + (p_{3}\times X) + (p_{4}\times Y), 
\label{eq:1}
\end{equation}
where $p_{N}$ is a free parameter, $T$ is the array of times, and $X$ and $Y$ are arrays of the positions of the trace on the detector, as well as an error inflation term that is added in quadrature to the measured errors. 

We fix the quadratic limb-darkening coefficients to the theoretical values computed with Set One of the MPS-ATLAS models with {\tt ExoTiC-LD} \citep{grantExoTiCLDThirtySeconds2024}, assuming the stellar parameters in Table~\ref{table:system_params}.
We also assume a circular orbit ($e=0$) and fix the orbital period of the planet to the value given in Table~\ref{table:system_params}. 
For our fits, we first run a Levenberg-Marquardt minimization on our combined astrophysical and instrumental noise model to obtain initial best-fit values to initialize 3$\times$ the number of walkers as free parameters for the Markov Chain Monte Carlo sampler {\tt emcee} \citep{DFM2013}. We run the fitter for 100,000 steps, discarding the initial 50,000 steps as burn-in. We take the median of the chains as the best-fit values and standard deviation of the chains as their associated uncertainties. We present the best-fit values from the white light curves in Table~\ref{table:retrieved-params}, the white light curves and their respective best-fit models in Fig.~\ref{fig:WLC_v1} and Fig.~\ref{fig:WLC_v2}, and the corresponding RMS as a function of bin size in Fig.~\ref{fig:allan_WLC}.

We utilize a similar method for fitting the spectroscopic light curves, with the exception of fixing the $i$, $a/R_{\star}$, $T_{0}$ values for each bin to the best-fit values from the white light curves of their respective detector and visit. We show the resulting transmission spectra in Fig.~\ref{fig:transmission_spectra}.

\section{Transmission spectrum} 
\label{SECTION-4:TRANMISSION_SPECTRUM}

We present the transmission spectra of \planet\ from the two independent reductions and two visits in the top panels of Fig.~\ref{fig:transmission_spectra}.
The combined spectra from the two visits are shown in the bottom panel of Fig.~\ref{fig:transmission_spectra}; for the \tiberius\ reduction, we combined the visits by taking the weighted mean of the individual visit spectra, and for the \eureka\ reduction the combined spectra results from a joint fit of the SLCs from each visit.
In Fig.~\ref{fig:spectral_precision}, we show the wavelength-dependent precisions achieved by the two reductions.
Neither reduction achieves the \texttt{pandexo} predicted precisions.
For visit 1, \tiberius\ and \eureka\ achieve an average transit depth precision of 25\,ppm and 24\,ppm in NRS1, and 45\,ppm and 40\,ppm in NRS2.
The precisions are similar for visit 2: 25\,ppm and 23\,ppm in NRS1, and 43\,ppm and 42\,ppm in NRS2.
The precisions scale as expected for stacked observations when combining the two visits: 18\,ppm and 17\,ppm in NRS1, and 33\,ppm and 29\,ppm in NRS2 (these correspond to 37\%, 28\% larger than \texttt{pandexo} predictions for NRS1, and 52\% and 33\% larger for NRS2).
These differences are comparable to those reported by other JWST/NIRSpec transmission spectroscopy studies that employed a similar number of groups per integration \citep[e.g.,][]{aldersonJWSTCOMPASSNIRSpec2024}.

In Appendix~\ref{APPENDIX-1:reductions}, Fig.~\ref{fig-app:transmission_spectra_visits}, we show the two reductions in the same panels for easier comparison; the spectra displayed there are identical to those in Fig.~\ref{fig:transmission_spectra}.
We see an average difference of 23\,ppm and 22\,ppm between \tiberius\ and \eureka\ in visit~1 and 2 respectively. 
These differences are below the average precisions of both pipelines, as given above.

The transmission spectra of \planet\ possess several interesting characteristics that are shared between the two reductions: (1) an offset between the NRS1 and NRS2 data; (2) a broad feature between 3 and 3.4\,$\mu$m in the NRS1 spectra; and (3) comparatively featureless NRS2 spectra, though a small peak at $\sim 4.7\,\mu$m may be present. Both the offsets and 3--3.4\,$\mu$m feature sizes are different between the two visits, while within one visit they are similar across the two reductions. 
Offsets between NRS1 and NRS2 transmission spectra have been seen before in NIRSpec/G395H observations \citep[e.g.,][]{wallackJWSTCOMPASSNIRSpec2024,moranHighTideRiptide2023,sarkar2024}, which have generally been attributed to instrumental rather than astrophysical effects.

\section{\texorpdfstring{Interpretation of the Transmission Spectrum of \planet}{Interpretation of the Transmission Spectrum of TOI-260b} }\label{SECTION-5:MODELLING}

\subsection{Non-Physical Models}
\label{sec:parametric}
In order to investigate the impact of the detector offset on our interpretation of the transmission spectra, as well as evaluate the significance of any features, we undertake similar non-physical modeling as in previous COMPASS studies and earlier works \citep[e.g.,][]{moranHighTideRiptide2023,scarsdaleJWSTCOMPASS352024,wallackJWSTCOMPASSNIRSpec2025}. In particular, we repeat the procedure and use the same custom code as in \citet{wallackJWSTCOMPASSNIRSpec2025} to fit non-physical models of various shapes to our data in a Bayesian retrieval framework. 
In brief, we consider five models: (1) a flat line, (2) a step function with an offset between NRS1 and NRS2, (3) a sloped line, (4) a single Gaussian in NRS1, and (5) a single Gaussian in NRS2. 
The Gaussian models are built atop the step function model and allow us to agnostically evaluate the significance of any deviations. 
We conduct the retrieval using \texttt{pymultinest} \citep{Buchner2014pymultinest}, which also computes the log of the Bayesian evidence, $\ln Z$, for each fit.
As standard in the literature, we employ the Bayes factor, $B$ ($\ln B = \Delta\ln Z$), to compare models, and adopt the qualitative thresholds given in Table~1 of \citet{thorngren2025}.
We apply this framework to both reductions of both visits, as well as the combined spectra. We consider wide uniform or log-uniform priors for all parameters and 1000 live points for each retrieval. 

\begin{figure*}
\centering
\includegraphics[width=\textwidth]{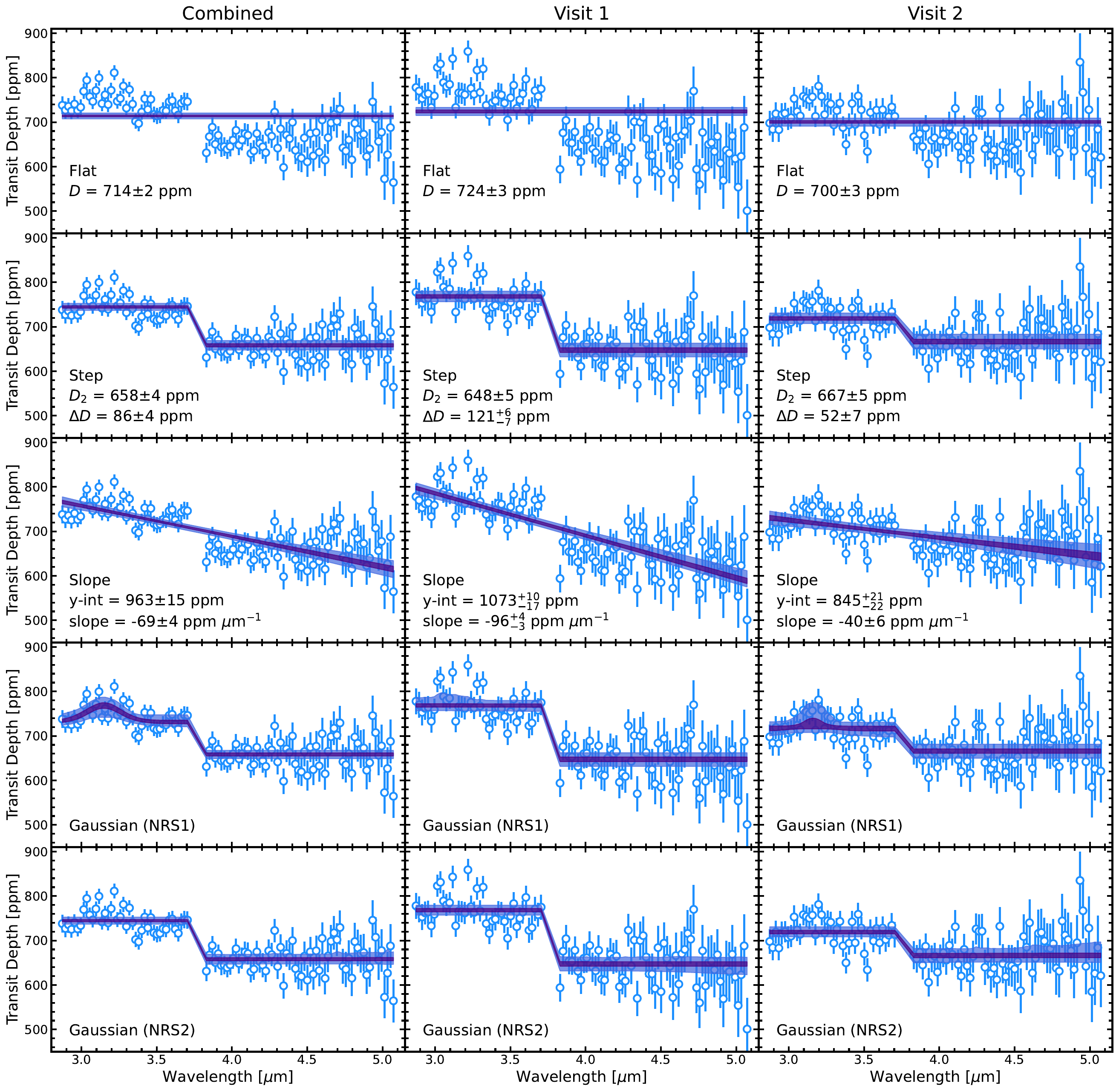}
\caption{Parametric model fits to the combined (left), visit 1 (middle), and visit 2 (right) \texttt{Tiberius} data reductions for the (top to bottom) flat, step function, slope, and NRS1 and NRS2 Gaussian models, respectively. For each panel, the data are given by the blue points, the 1$\sigma$ bounds of the best fit models are given by the inner, darker blue bands, and the 3$\sigma$ bounds are given by the outer, lighter blue bands. For the flat, step function, and slope models the best fit parameter values and associated uncertainties are given in the figure, where $D$ is the transit depth, $D_2$ is the NRS2 transit depth, and $\Delta D$ is the offset between the NRS1 and NRS2 detectors; those of the Gaussian models are given in Table \ref{tab:gausstab}.}
\label{fig:unphysmod_tiberius}
\end{figure*}

\begin{figure*}
\centering
\includegraphics[width=\textwidth]{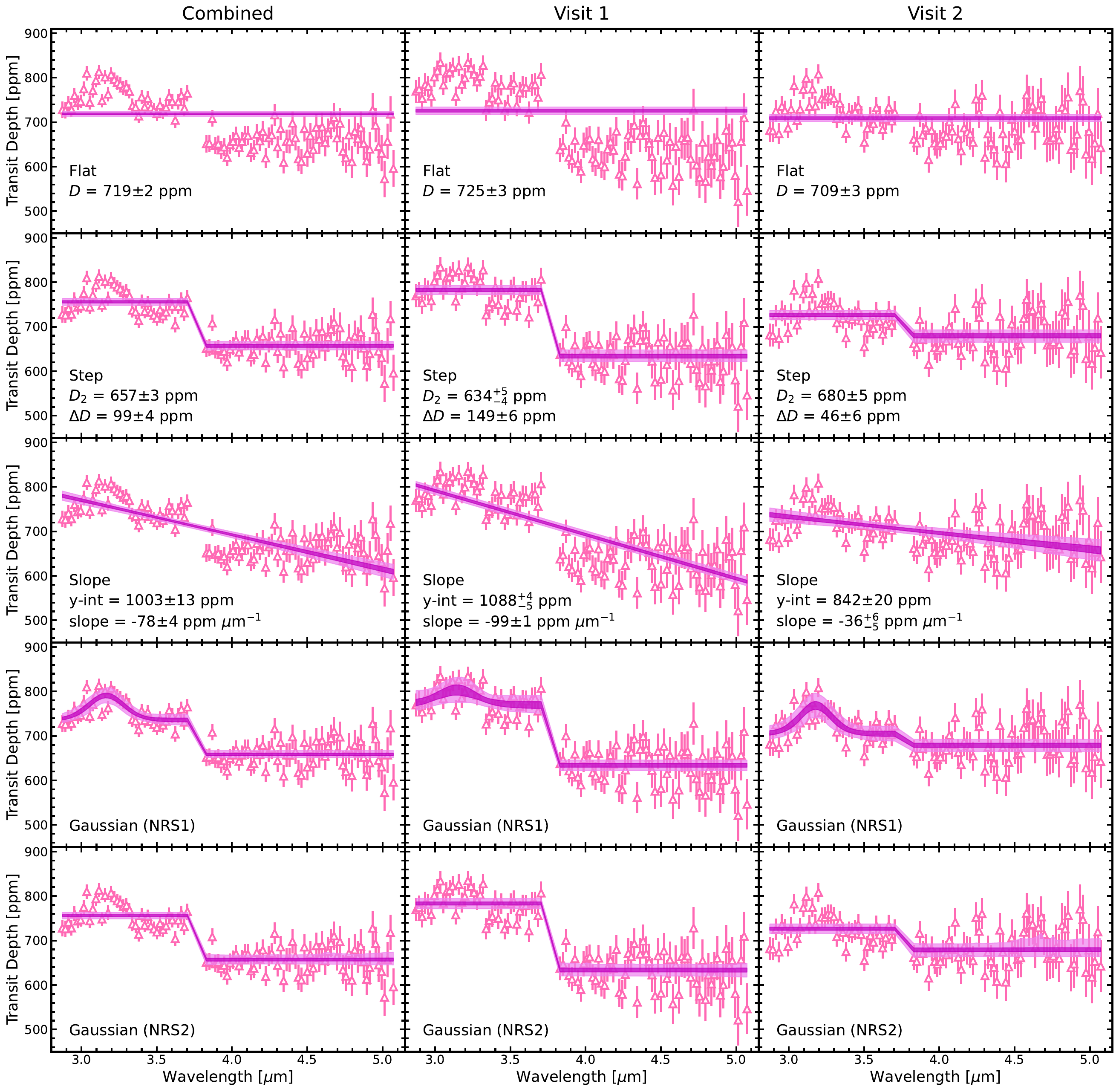}
\caption{Same as Fig. \ref{fig:unphysmod_tiberius}, but for the \texttt{Eureka!} reductions. The data are given by the pink points, the 1$\sigma$ bounds of the best fit models are given by the inner, darker magenta bands, and the 3$\sigma$ bounds are given by the outer, lighter magenta bands. }
\label{fig:unphysmod_eureka}
\end{figure*}

We present the retrieval results and best-fit parameter values for each case in Fig.~\ref{fig:unphysmod_tiberius} and Fig.~\ref{fig:unphysmod_eureka} and Table~\ref{tab:gausstab}, and the $\ln Z$ values in Table~\ref{tab:lnz}. 
These show that the step function model is significantly preferred \citep[$\Delta \ln Z>5$;][]{thorngren2025} over the flat line and slope models for all visits and reductions.
In addition, the retrieved detector offsets are much greater in visit 1 compared to visit 2. For example, for the \tiberius\ reduction of visit 1 we find $\Delta D=121^{+6}_{-7}$\,ppm, while for visit 2 $\Delta D=52\pm7$\,ppm; similarly, for the \eureka\ reduction of visit 1 we find $\Delta D=149\pm6$\,ppm, while for visit 2 $\Delta D=46\pm6$\,ppm.
Differences in detector offsets between visits have been seen in other exoplanet NIRSpec G395H datasets \citep[e.g.,][]{aldersonJWSTCOMPASSNIRSpec2025}; we do not correct for these offsets, and instead leave them as a free parameter when using atmospheric retrievals to place constraints on the atmospheric composition of \planet (\S\ref{sec:platon}).

\begin{table*}
\centering
\caption{Best fit parameters (mean and 1$\sigma$ uncertainty) of the NRS1 and NRS2 single Gaussian models for the \texttt{Tiberius} (\texttt{T}) and \texttt{Eureka!} (\texttt{E!}) combined (c), visit 1 (v1) and visit 2 (v2) reductions.}
\begin{tabular}{lcccccc}
\toprule
Parameter & \texttt{T} (c) & \texttt{T} (v1) & \texttt{T} (v2) & \texttt{E!} (c) & \texttt{E!} (v1) & \texttt{E!} (v2) \\
\midrule
\multicolumn{7}{c}{Gaussian (NRS1)} \\
\midrule
Amplitude (ppm) & 39$^{+7}_{-8}$ & 10$^{+36}_{-8}$ & 15$^{+32}_{-13}$ & 56$^{+6}_{-7}$ & 41$^{+11}_{-14}$ & 65$\pm$10 \\
Central wavelength ($\mu$m) & 3.15$\pm$0.02 & 5.68$^{+2.95}_{-3.5}$ & 3.2$^{+4.54}_{-1.44}$ & 3.16$^{+0.02}_{-0.01}$ & 3.14$\pm$0.04 & 3.18$\pm$0.02 \\
Standard deviation ($\mu$m) & 0.16$\pm$0.03 & 0.07$^{+1.09}_{-0.06}$ & 0.1$^{+0.83}_{-0.09}$ & 0.17$\pm$0.02 & 0.18$^{+0.07}_{-0.04}$ & 0.15$\pm$0.03 \\
NRS1 transit depth (ppm) & 731$\pm$4 & 768$\pm$4 & 717$^{+5}_{-7}$ & 736$\pm$4 & 768$^{+9}_{-6}$ & 705$\pm$6 \\
NRS2 transit depth (ppm) & 658$\pm$3 & 648$\pm$5 & 666$\pm$5 & 658$\pm$3 & 634$\pm$4 & 679$\pm$5 \\
\midrule
\multicolumn{7}{c}{Gaussian (NRS2)} \\
\midrule
Amplitude (ppm) & 9$^{+30}_{-7}$ & 10$^{+34}_{-8}$ & 8$^{+28}_{-6}$ & 8$^{+31}_{-6}$ & 8$^{+32}_{-6}$ & 8$^{+34}_{-6}$ \\
Central wavelength ($\mu$m) & 5.61$^{+2.98}_{-3.91}$ & 4.96$^{+3.37}_{-3.14}$ & 5.48$^{+3.04}_{-3.66}$ & 5.55$^{+2.87}_{-3.76}$ & 5.08$^{+3.33}_{-3.53}$ & 5.17$^{+3.15}_{-3.52}$ \\
Standard deviation ($\mu$m) & 0.06$^{+1.5}_{-0.06}$ & 0.06$^{+1.36}_{-0.05}$ & 0.08$^{+1.51}_{-0.08}$ & 0.1$^{+1.73}_{-0.09}$ & 0.1$^{+1.9}_{-0.09}$ & 0.05$^{+1.52}_{-0.05}$ \\
NRS1 transit depth (ppm) & 745$^{+3}_{-2}$ & 768$\pm$4 & 719$\pm$4 & 756$\pm$2 & 783$\pm$3 & 727$\pm$3 \\
NRS2 transit depth (ppm) & 658$\pm$4 & 647$^{+5}_{-6}$ & 666$^{+5}_{-6}$ & 657$^{+3}_{-4}$ & 633$^{+4}_{-5}$ & 679$^{+4}_{-5}$ \\
\bottomrule
\end{tabular}
\label{tab:gausstab}
\end{table*}

\begin{table*}
\centering
\caption{$\Delta \ln{Z}$ between the parametric models and the step function models for the \texttt{Tiberius} (\texttt{T}) and \texttt{Eureka!} (\texttt{E!}) combined (c), visit 1 (v1) and visit 2 (v2) reductions. Negative $\Delta \ln{Z}$ indicates lower preference compared to the step function model.}
\begin{tabular}{lcccccc}
\toprule
Model & \texttt{T} (c) & \texttt{T} (v1) & \texttt{T} (v2) & \texttt{E!} (c) & \texttt{E!} (v1) & \texttt{E!} (v2) \\
\midrule
Flat line & -174 & -166 & -28 & -280 & -326 & -30 \\
Step function & 0 & 0 & 0 & 0 & 0 & 0 \\
Slope & -33 & -23 & -10 & -52 & -64 & -10 \\
Gaussian (NRS1) & 5 & 0 & 0 & 21 & 3 & 11 \\
Gaussian (NRS2) & 0 & 1 & 0 & 0 & 1 & -1 \\
\bottomrule
\end{tabular}
\label{tab:lnz}
\end{table*}

For the Gaussian models, we find that the NRS1 Gaussian is strongly preferred ($\Delta \ln Z>5$) over the step function for the combined and visit 2 spectra from the \texttt{Eureka!}\ reduction, weakly preferred ($\Delta \ln Z\sim3-5$) for the combined spectrum from the \texttt{Tiberius} reduction and the visit 1 spectrum of the \texttt{Eureka!} reduction, and not preferred for either of the individual visit \texttt{Tiberius} spectra ($\Delta \ln Z\sim0$). 
For the spectra where the NRS1 Gaussian is strongly preferred over the step function, we find a consistent feature centered at $\sim$3.17 $\mu$m with a 1$\sigma$ width of $\sim$0.16 $\mu$m (Table \ref{tab:gausstab}). 
However, the amplitude of the feature varies significantly between the reductions, with a $\sim$3$\sigma$ difference between those of the combined \texttt{Tiberius} and \texttt{Eureka!}\ spectra. 
We will attempt to interpret the 3.17 $\mu$m feature in \S\ref{SECTION-6:DISCUSSION}. 
In contrast, the NRS2 Gaussian model is not preferred over the step function model, with $\Delta \ln Z\leq1$ for all reductions and visits, suggesting a lack of any significant features in the NRS2 spectra. This is further supported by the retrieved Gaussian parameter values in Table \ref{tab:gausstab}, which exhibit large uncertainties. 

To summarize, the transmission spectrum of \planet\ is most aptly described by a step function with the step occurring between the NRS1 and NRS2 detectors. In addition, there may be evidence towards a potential, broad absorption feature centered at 3.17 $\mu$m. 
In the next section, we explore what atmospheric thermochemical equilibrium models can tell us about the atmospheric composition, and whether we are sensitive to an aerosol layer and/or the solid surface of the planet. 

\subsection{PLATON Retrievals}\label{sec:platon}

Thermochemical equilibrium models can be helpful in giving estimates of the atmospheric composition when the spectrum is mostly featureless. 
In our case, they can also inform our interpretation of the 3.17 $\mu$m feature. 
As with previous studies in the COMPASS program \citep[e.g.,][]{TeskeJWSTCOMPASS2025,wallackJWSTCOMPASSNIRSpec2025}, we consider thermochemical equilibrium Bayesian retrievals to place constraints on the atmospheric metallicity/mean molecular weight and the presence of an aerosol layer and/or the solid planetary surface.

As in \citet{wallackJWSTCOMPASSNIRSpec2025}, we use the PLATON \citep[PLanetary Atmospheric Tool for Observer Noobs, version 6;][]{Zhang2025PLATON} retrieval code for this study with the default R=20,000 resampled opacities and the \texttt{pymultinest} sampler with 1,000 live points.
Our likelihood function takes the form,
\begin{equation}
    \ln \mathcal{L}(y|\mathbf{x}) = -\frac{1}{2}\sum_{i=1}^N \Big[ \frac{(y_i - F_i(\mathbf{x}))^2}{\sigma_i^2} +\ln(2 \pi\sigma_i^2) \Big]
\end{equation}
where $y_i$ is the data, $\sigma_i$ the data uncertainties, and $F_i(x)$ the forward-modeled transmission spectrum for a given set of input parameters.
Table \ref{tab:platonparams} gives our retrieval parameters, $\mathbf{x}$, and their priors, where we have converted the measured stellar radius $R_*$ and planet mass $M_p$ from Table \ref{table:system_params} into Gaussian priors, and assigned uniform priors for all other parameters. 
For $M_p$ we assume a symmetric Gaussian prior using the larger of the inferred uncertainties to be conservative.
We assume a wide uniform prior for the planet radius $R_p$ around the measured value to account for the unknown transit pressure.
We assume a fixed isothermal temperature profile at the equilibrium temperature of \planet\ (489\,K) and an effective temperature for the star of 4026\,K (Table \ref{table:system_params}). 
The C/O ratio of the atmosphere is fixed to solar \citep[0.59;][]{Asplund2021}, since variations thereof do not significantly impact the atmospheric mean molecular weight at the temperature of \planet \citep{teskeJWSTCOMPASSNIRSpec2025}.

\begin{figure*}
\centering
\includegraphics[width=\textwidth]{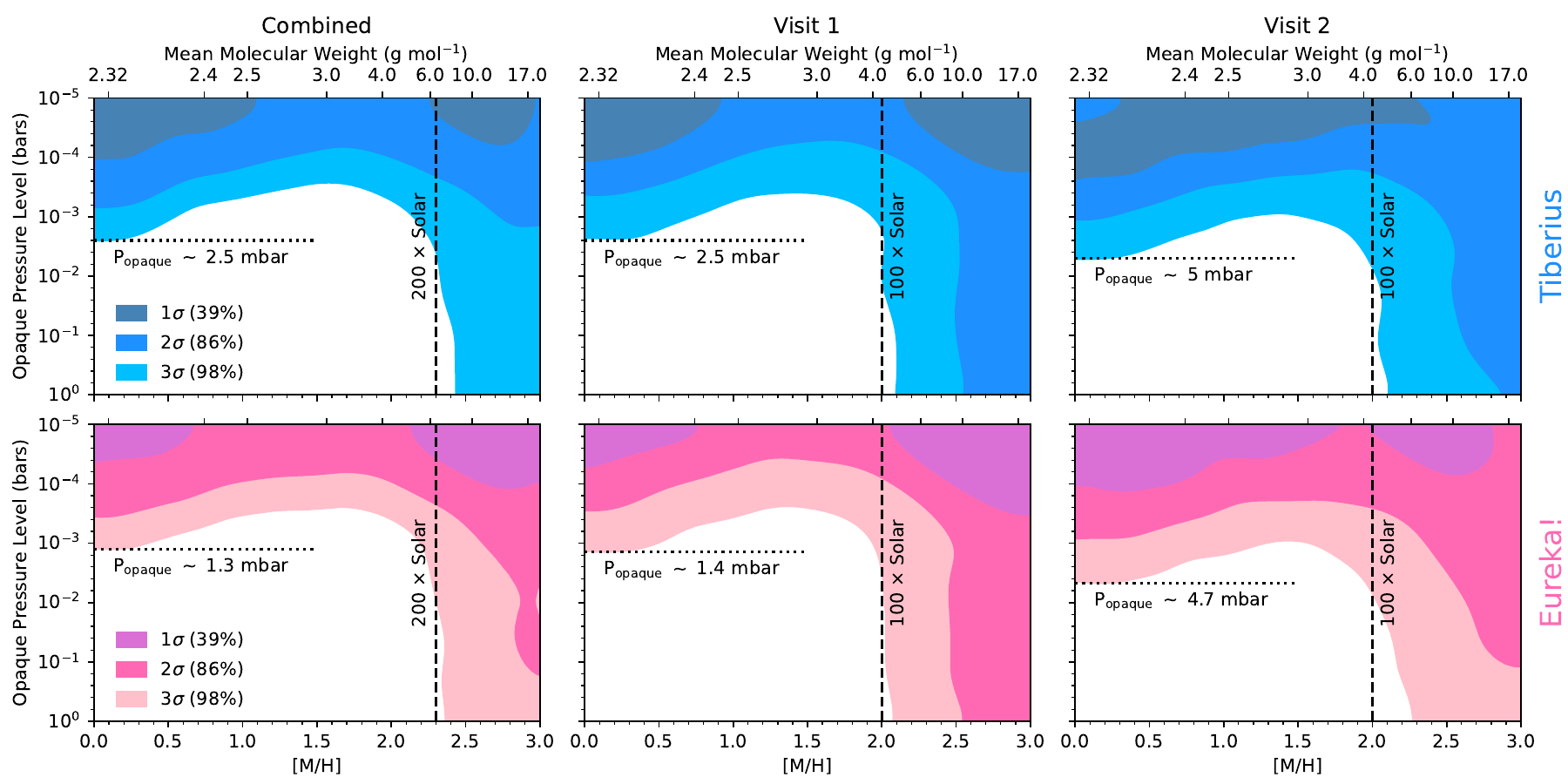}
\caption{$\sigma$-rejection contours in metallicity-opaque pressure level ($P_{\rm opaque}$) space from our PLATON retrievals for \texttt{Tiberius} (top) and \texttt{Eureka!}\ (bottom) reductions of the combined (left), visit 1 (middle), and visit 2 (right) data. The 1, 2, and 3$\sigma$ contours in 2D are indicated with increasingly lighter color shades. The mean molecular weight at 1 bar in an isothermal atmosphere with \planet's equilibrium temperature is shown in the top axes and corresponds to the metallicities at the bottom axes, computed using PLATON assuming C/O = 0.59 and thermochemical equilibrium. Minimum allowed clear-atmosphere metallicities are indicated by the vertical dashed lines, while the maximum allowed P$_{\rm opaque}$ values at lower metallicities are indicated by the horizontal dotted lines.}
\label{fig:metpcld}
\vspace{2em}
\end{figure*}

\begin{table}
\centering
\begin{threeparttable}
    \caption{Free parameters and priors in our PLATON retrievals.}
    \begin{tabular}{ll}
    \toprule
        Parameter & Prior \tnote{$\dagger$} \\
        \midrule
         $R_{*}$ ($R_{\odot}$)  & $\mathcal{N}(0.618, 0.06)$\\
         $M_\mathrm{p}$ ($M_{\earth}$)  & $\mathcal{N}(3.97,2.39)$\\
         $R_\mathrm{p}$ ($R_{\earth}$)  & $\mathcal{U}[1,2.5]$\\
         log[M/H ($\times$ Solar)]   & $\mathcal{U}[0,3]$ \\
         log[$P_{opaque}$ (bars)]  &  $\mathcal{U}[-5,0]$ \\
         Offset, $\Delta D$ (ppm)  & $\mathcal{U}[-500,500]$\\
         \bottomrule
    \end{tabular}
     \begin{tablenotes}
       \item [$\dagger$] For Gaussian priors this gives the mean and standard deviation. For uniform priors this gives the lower and upper bounds. 
     \end{tablenotes}
    \label{tab:platonparams}
\end{threeparttable}
\end{table}

\begin{figure}
\centering
\includegraphics[width=0.45\textwidth]{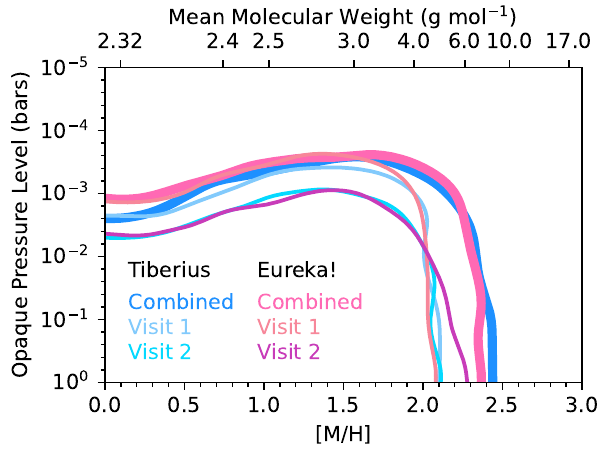}
\caption{3$\sigma$-rejection contours in metallicity-opaque pressure level space from our PLATON retrievals from all reductions and visits shown in Fig.~\ref{fig:metpcld}, over-plotted for ease of comparison. The contours for the combined datasets are shown in the thicker curves. The parameter space below and to the left of each curve is excluded by the corresponding dataset at $>3\sigma$.}
\label{fig:metpcldcombined}
\end{figure}

\begin{figure*}
\centering
\includegraphics[width=\textwidth]{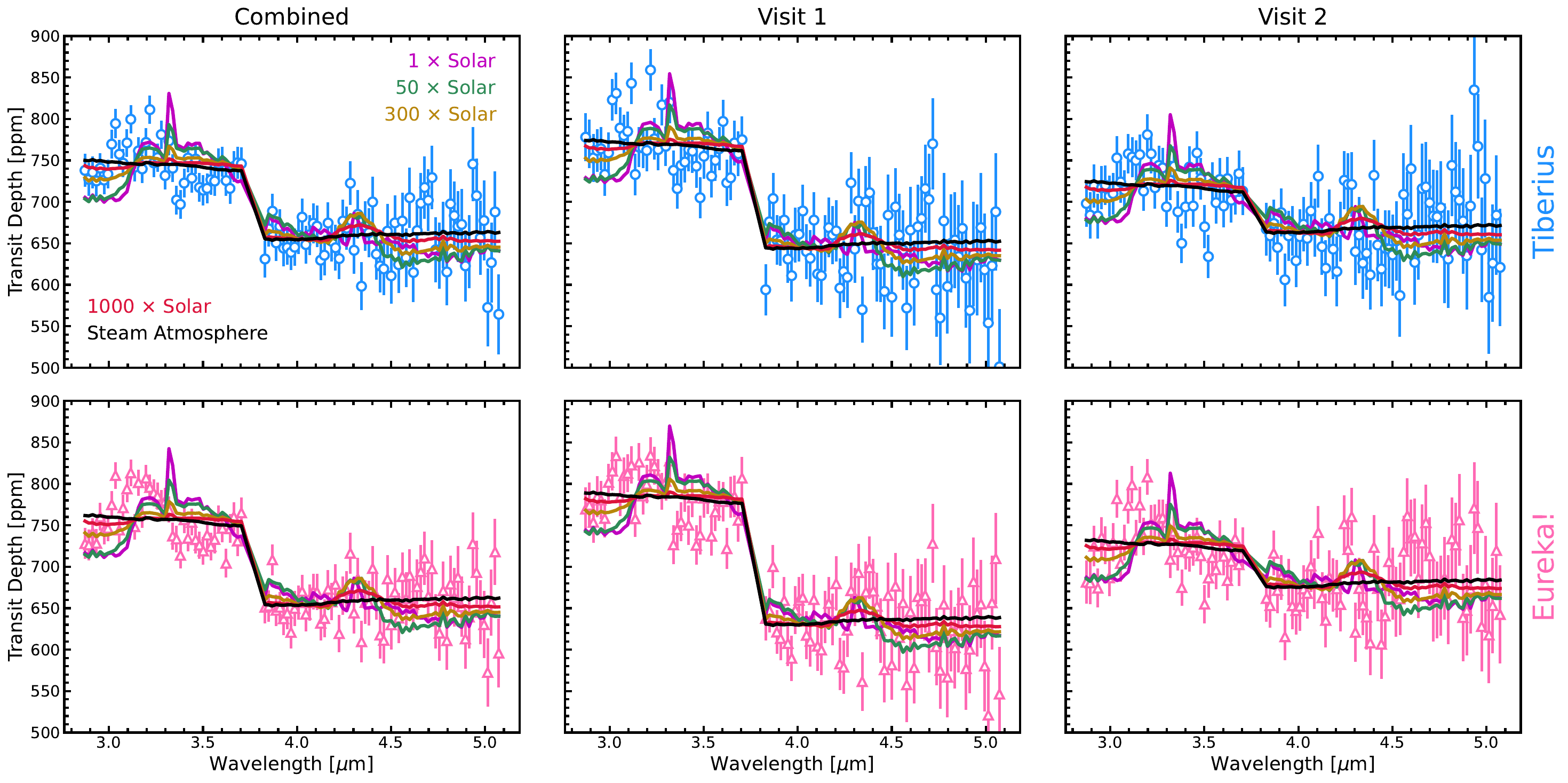}
\caption{Comparisons between representative clear-sky thermochemical equilibrium atmospheric models (solar metallicity: magenta; 50 $\times$ solar: green; 300 $\times$ solar: yellow; 1000 $\times$ solar: red; pure H$_2$O atmosphere: black) and the \texttt{Tiberius} (top) and \texttt{Eureka!}\ (bottom) reductions of the combined (left), visit 1 (middle), and visit 2 (right) data. The reduced chi-squared values of each model compared to the data are given in Table \ref{tab:redchisqr}.  } 
\label{fig:models}
\end{figure*}

\begin{table*}
\centering
\caption{$\chi^2/N$ values of the thermochemical equilibrium model fits for the \texttt{Tiberius} (\texttt{T}) and \texttt{Eureka!}\ (\texttt{E!})  combined (c), visit 1 (v1) and visit 2 (v2) reductions shown in Figure \ref{fig:models}.}
\begin{tabular}{lcccccc}
\toprule
Model & \texttt{T} (c) & \texttt{T} (v1) & \texttt{T} (v2) & \texttt{E!} (c) & \texttt{E!} (v1) & \texttt{E!} (v2)  \\
\midrule
1 $\times$ solar & 2.81 & 2.01 & 1.72 & 3.4 & 2.21 & 2.26 \\
50 $\times$ solar & 2.43 & 1.84 & 1.53 & 3.0 & 2.0 & 2.07 \\
300 $\times$ solar & 1.63 & 1.41 & 1.17 & 2.07 & 1.5 & 1.64 \\
1000 $\times$ solar & 1.42 & 1.3 & 1.08 & 1.82 & 1.37 & 1.53 \\
Steam atmosphere & 1.34 & 1.26 & 1.04 & 1.71 & 1.34 & 1.46 \\
\bottomrule
\end{tabular}
\label{tab:redchisqr}
\end{table*}

Figures \ref{fig:metpcld} and \ref{fig:metpcldcombined} show the constraints imposed on the atmosphere of \planet with the PLATON retrievals.
As with previous results from the COMPASS program, we find that \planet's largely featureless transmission spectra lead to a preference for high atmospheric metallicity models and/or high altitude aerosol models. 
In particular, the combined spectra from both reductions are able to rule out to $>$3$\sigma$ metallicities $\leq$200 $\times$ solar (atmospheric mean molecular weight MMW $\leq6$ g mol$^{-1}$). The constraints from the individual visits are looser, ruling out metallicities $\leq$100 $\times$ solar (MMW $\leq4$ g mol$^{-1}$). All of these metallicity constraints hold for opaque pressures greater than a few mbar, while at lower opaque pressures lower metallicities are allowed, including down to solar metallicity.

In Figure \ref{fig:models}, we compare several representative clear atmosphere models to the data, which demonstrate why we can generally rule out low metallicity scenarios when the opaque pressure level is deep.
These models are computed at various metallicities using PLATON assuming thermochemical equilibrium at the equilibrium temperature of \planet and solar C/O \citep[0.59;][]{Asplund2021}. 
We also include a pure H$_2$O steam model as an example of an outgassed, secondary atmosphere. 
These models are fit to the data using \texttt{scipy.minimize} with the NRS2 transit depth and an offset between NRS1 and NRS2 as free parameters; the $\chi^2/N$ values (chi-squared divided by the total number of data points) are reported in Table \ref{tab:redchisqr}. Higher metallicity models exhibit lower $\chi^2/N$ values, as their spectral features--consisting mostly of the CH$_4$ and CO$_2$ features at 3.3 and 4.3 $\mu$m, respectively--become more muted due to the decreasing scale height resulting from the increasing atmospheric mean molecular weight. The steam atmosphere model also exhibits low $\chi^2/N$ values due to its small scale height, and also the lack of high amplitude water features at 3-5 $\mu$m. In other words, our data largely prefer flat and featureless spectral models, which is enabled by increasing the metallicity or adding a high altitude aerosol layer.  

Interestingly, no spectral features from our models fit the 3.17 $\mu$m feature in the NRS1 data, with the CH$_4$ absorption band at 3.3 $\mu$m in the solar metallicity model coming the closest in feature amplitude and wavelength. As such, if the 3.17 $\mu$m feature originates from gas absorption in the atmosphere, then the atmospheric scale height cannot be much smaller than that at solar metallicity, while elemental ratios like C/O must be significantly non-solar to produce gases that can absorb at 3.17 $\mu$m not seen in our solar C/O models. 
We explore this further in the next section (\S\ref{section-6:317_astro}).

\section{Discussion} 
\label{SECTION-6:DISCUSSION}

\subsection{\texorpdfstring{The Origins of the 3.17\,$\mu$m Feature}{The Origins of the 3.17um Feature}}
\label{section-6:3_18_feature}
\subsubsection{Systematics origin}
The 3.17\,$\mu$m feature is seen across several of our reductions and visits, necessitating a closer look at its origins. 
One possibility is that it is a manifestation of data systematics, as might be suggested by the varying feature amplitudes and detection significances across the different reductions and visits (\S\ref{sec:parametric}).

\begin{figure}
    \centering
    \includegraphics[width=\columnwidth]{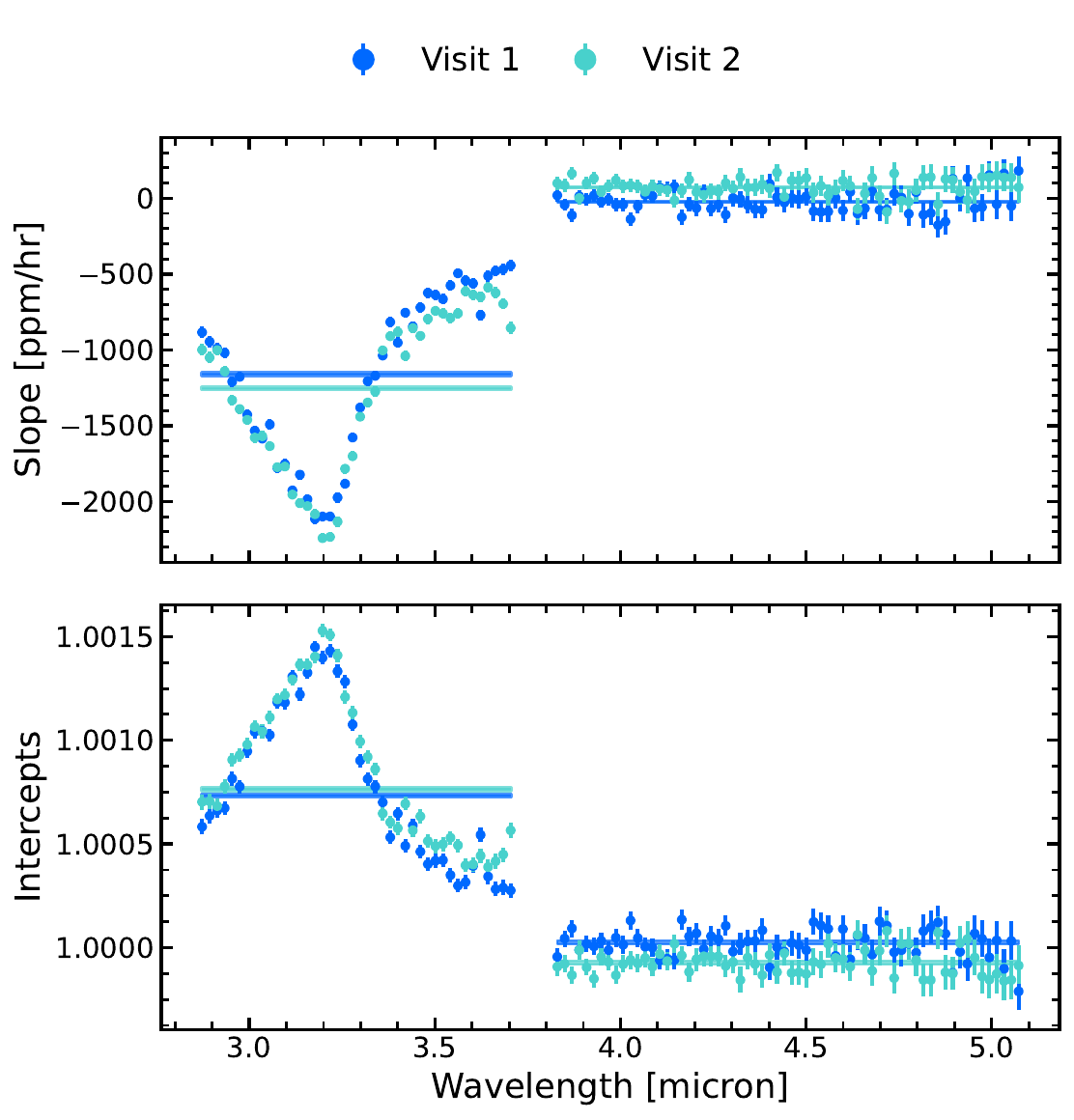}
    \caption{The fitted coefficients for the systematics models fitted to the \texttt{Tiberius} spectroscopic light curves.
    The values from the broadband light curve fits (Fig.~\ref{fig:WLC_v1} and Fig.~\ref{fig:WLC_v2}) are shown as horizontal bars for reference.}
    \label{fig-app:tiberius_slopes_intercepts}
\end{figure}

Upon investigation, we noted a characteristic trend in the fitted systematics model parameters.
In both reductions described in \S\ref{SECTION-3:REDUCTION}, we fit linear polynomials to the NRS1 and NRS2 spectroscopic light curves; such linear trends in time are standard in NIRSpec light curves \citep[see e.g.,][]{espinozaSpectroscopicTimeseriesPerformance2022}.
We show the fitted linear coefficients of the \tiberius\ spectroscopic systematic models in Fig.~\ref{fig-app:tiberius_slopes_intercepts}, along with the WLC values.
Though reasonably consistent across the NRS2 bins, the exposure-long slopes vary across NRS1, peaking near the 3.17\,$\mu$m region.
Such wavelength dependence has been seen in other NIRSpec G395H transit data, including other COMPASS targets.
This might be suggestive of a wavelength-dependent detector systematic: if this is manifesting in the transmission spectrum, it may be that there exist further underlying systematics that are similarly correlated but are not yet fully captured by standard metrics.
Other than one outlier bin corresponding to $3.197$\,$\mu$m in visit 1, the noise properties of the bins in this region are not significantly different from the rest of the NRS1 bandpass, and the residuals display little red noise (see Appendix Fig.~\ref{fig-app:allan_SLC}).
Nevertheless, we explored adding dimensions to our \tiberius\ SLC systematics models.

We assess the other characteristic detector trace parameters output by \tiberius\, including the \texttt{ypositions} (global shift of the spectrum along the dispersion direction), \texttt{xpositions} (the smoothed trace center pixel positions in the cross-dispersion direction, i.e., the position of the aperture trace), and \texttt{sky} (the median of the `background' pixels that were used for the 1/f correction, in each column).
The inclusion of these time arrays as additional dimensions in the systematics models made no observable difference to the transmission spectrum, and the retrieved parameters were in agreement with the simpler models (example provided in Appendix~\ref{APPENDIX-1:reductions}, Table~\ref{table-app:WLC_params}).

\begin{figure}
    \centering
    \includegraphics[width=\linewidth]{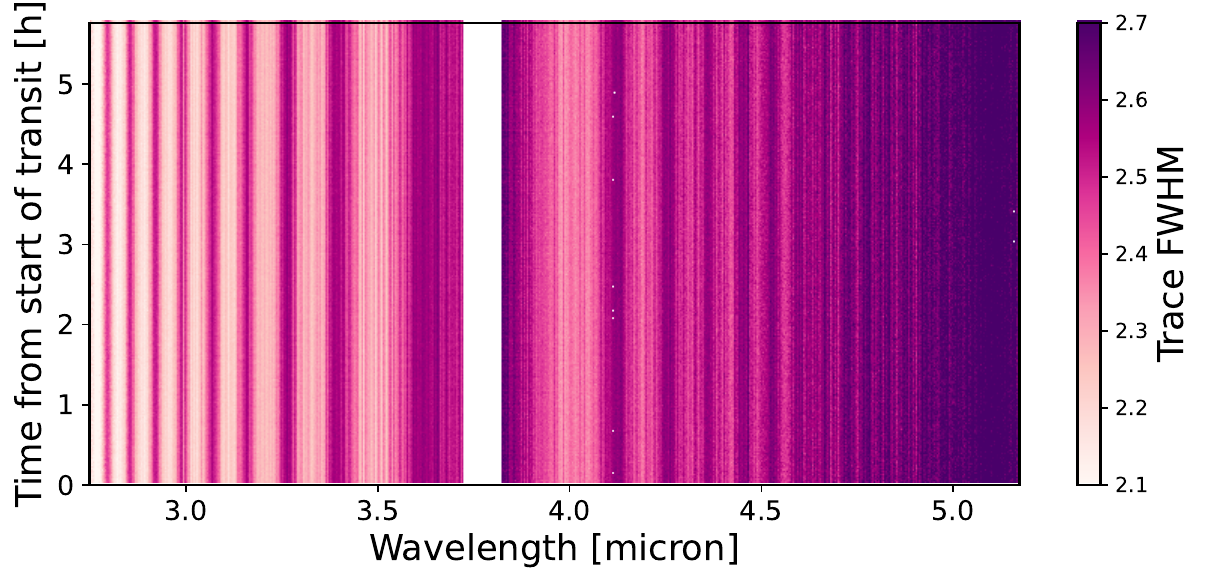}
    \caption{The fitted FWHM of the aperture trace on detectors NRS1 and NRS2, using \texttt{Tiberius} on visit 1.}
    \label{fig-app:fwhm}
\end{figure}

The measured trace \texttt{fwhm} (the full width half maximum of the Gaussians fitted along the cross-dispersion direction, i.e. the width of the aperture trace) exhibits sinusoidal behaviour along the dispersion axis; we show the measured \tiberius\ \texttt{fwhm} for each integration from visit 1 in Fig.~\ref{fig-app:fwhm}.
This behavior was reported in \citet{espinozaSpectroscopicTimeseriesPerformance2022}, who measured a sinusoid amplitude of 0.6 pixels.
They attributed this to a measurement effect, rather than a true representation of the detector/optics; since the trace is tilted, the PSF lands on different regions of each pixel, thereby causing a fluctuation in the measured FWHM. 
To investigate if adding in some metric for the variability of the aperture trace has an impact on the transmission spectrum, we introduce the 2D \texttt{fwhm} array (Fig.~\ref{fig-app:fwhm}) as a detrending parameter for the SLC systematics models, which take the form
\begin{equation}
    S_n(t) = a_0 + a_1 t + a_2 F_n (t).
\end{equation}
We first smooth the \texttt{fwhm} by fitting a fourth order polynomial to the minima of the sinusoidal \texttt{fwhm} for each integration, similarly to \citet{espinozaSpectroscopicTimeseriesPerformance2022}.
For each wavelength bin, $n$, we average the smoothed \texttt{fwhm} across the binned pixels to produce a 1D array $F_n (t)$.
We fixed the limb-darkening coefficients to that adopted for the original SLC fitting, and fix $a/R_*$, $i$, and $T_\mathrm{mid}$ to the values from the original WLC fitting (Table~\ref{table:retrieved-params}).
Fitting the same, 60\,min-trimmed SLCs, we found that the inclusion of the wavelength-dependent \texttt{fwhm} had no observable impact on the RMS of the binned residuals, and little-to-no difference on the transmission spectrum (as an example, there was an average difference of 4ppm for the visit 1 NRS1 transmission spectrum).\footnote{We note that we separately investigated the use of the oversampling factor in the \tiberius\ spectral extraction of NRS1, whereby the profile of each pixel column is linearly interpolated onto a finer grid. Using an oversampling factor of 10 in the cross-dispersion direction,  we saw no improvement in the RMS of the white light curve, no change in the fitted astrophysical parameters, and only very minor differences in the transmission spectrum (average of 5\,ppm) and precisions.}
Ultimately, the inclusion of additional linear terms in our systematics models had no impact on the shape of the recovered NRS1 transmission spectrum.

We also investigated the inclusion of engineering data from the JWST Engineering Mnemonic
Database, as has been done for previous NIRSpec G395H observations that show potentially non-physical results \citep{wallackJWSTCOMPASSNIRSpec2024}. 
Exploring correlations with engineering data did not elucidate the origin of this feature.
This is not surprising, as the engineering mnemonic exploration is specifically optimized to reduce the amount of red noise, not wavelength-correlated noise, and the specific spectral bins around the feature do not show excessive red noise.

While we do not rule out a systematics origin of the feature seen in the transmission spectrum of \planet, we have shown that it is robust against different detrending methods that are currently standard for JWST/NIRSpec G395H observations.
For further exploration of systematics in the COMPASS JWST/NIRSpec observations, we refer the reader to \citet{gordon2025}.

\begin{table}
\centering
\begin{threeparttable}
    \caption{$\Delta \ln{Z}$ between the full free retrieval with PLATON and those with certain chemical species removed. Negative $\Delta \ln{Z}$ indicates lower preference. }
    \begin{tabular}{lcc}
    \toprule
        Retrieval Setup & \texttt{Tiberius} & \texttt{Eureka!} \\ 
        \midrule
         All Species & 0 & 0 \\
         No C$_2$H$_2$, C$_2$H$_4$ & -0.25 & -11.67 \\
         No C$_2$H$_2$, C$_2$H$_4$, C$_3$ & -3.73 & -14.28 \\
         No C$_2$H$_2$, C$_2$H$_4$, C$_3$, NH$_3$  & -6.47 & -21.30 \\
         Common Species Only\tnote{$\star$} & -8.16 & -21.65 \\
         \bottomrule
    \end{tabular}
     \begin{tablenotes}
       \item [$\star$] H$_2$O, CH$_4$, CO$_2$, CO, SO$_2$, and H$_2$S 
     \end{tablenotes}
    \label{tab:freechem}
\end{threeparttable}
\end{table}

\begin{figure*}
    \centering
    \includegraphics[width=0.79\linewidth]{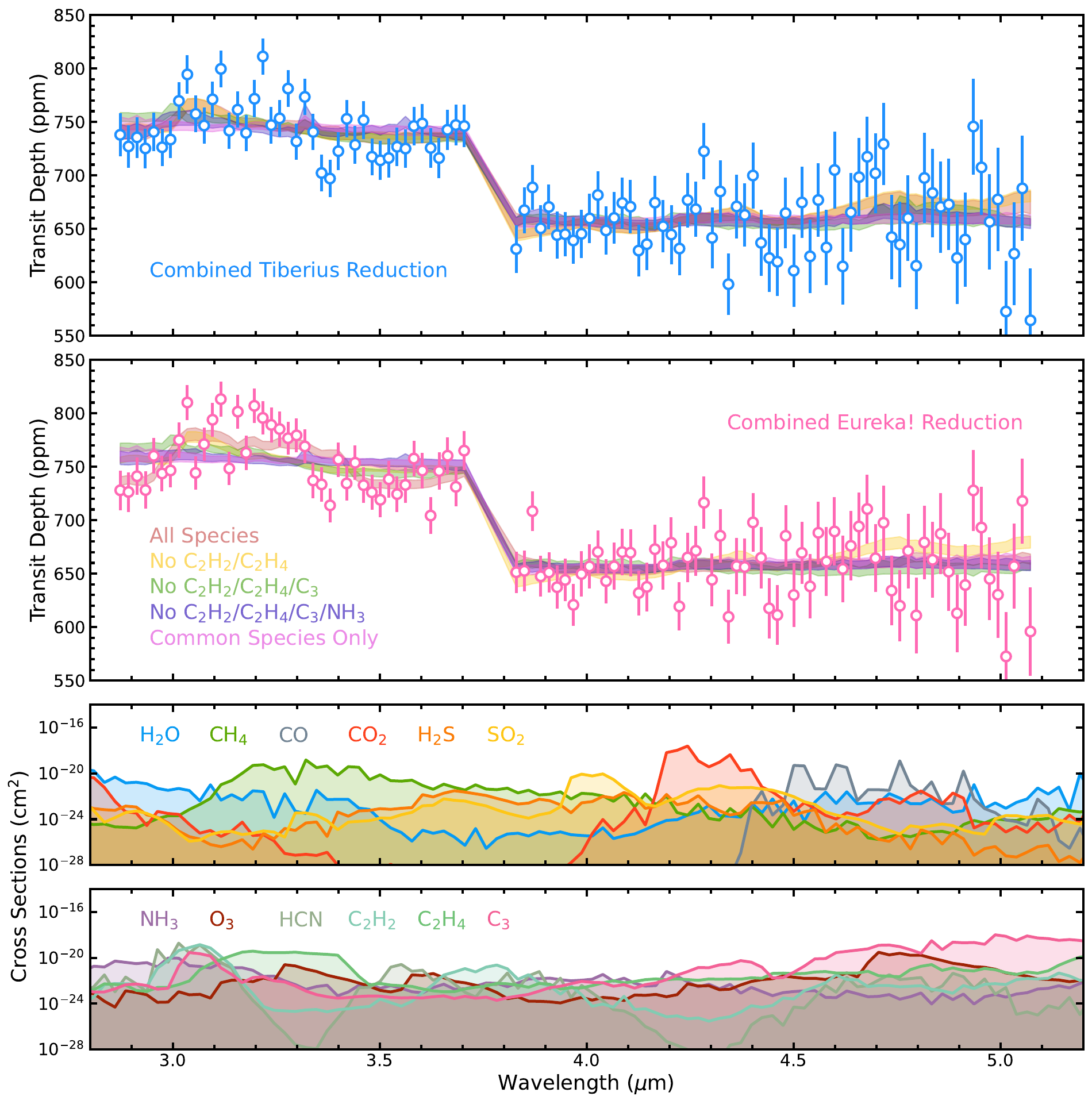}
    \caption{Free retrieval results (top two panels) and cross sections of considered chemical species (bottom two panels) from our investigation of the 3.17\,$\mu$m feature. Free retrievals were conducted on the combined \texttt{Tiberius} (top, blue) and \texttt{Eureka!} (middle-top, pink) reductions. The 1$\sigma$ range in retrieved spectra are shown in the colored bands for all considered species (red), all species except C$_2$H$_2$ and C$_2$H$_4$ (yellow), all species except for C$_2$H$_2$, C$_2$H$_4$, and C$_3$ (green), all species except for C$_2$H$_2$, C$_2$H$_4$, C$_3$, and NH$_3$ (dark blue), and only the common species (H$_2$O, CH$_4$, CO$_2$, CO, SO$_2$, and H$_2$S; light pink). }
    \label{fig-app:frspec}
\end{figure*}

\subsubsection{Astrophysical origin}
\label{section-6:317_astro}
Aside from systematics, it is also possible that the 3.17\,$\mu$m feature originates from gas absorption in \planet's atmosphere. 
As discussed in \S\ref{sec:platon} and shown in Fig.~\ref{fig:models}, the amplitude of the feature suggests an atmospheric mean molecular weight not that much greater than that at solar metallicity, as otherwise the spectral features would be highly muted. In addition, gases produced from thermochemical equilibrium assuming solar C/O cannot explain the feature. We therefore consider absorption by chemical species that can result from atmospheres with significantly non-solar elemental ratios for metals and/or are impacted by chemical disequilibrium processes. 
In addition, recent theoretical work has predicted that helium-dominated atmospheres may arise on sub-Neptunes with similar masses and radii as \planet, due to mass fractionation during atmospheric loss \citep{Malsky2023,Cherubim2025}. The relatively low H abundance in such atmospheres would cause the formation of chemical species often found in high metallicity atmospheres, while the He background would ensure a scale height within a factor of 2 of that of a H$_2$/He-dominated atmosphere. Under these considerations, we perform a non-exhaustive search of the DACE\footnote{\url{https://dace.unige.ch}} and ExoMol\footnote{\url{https://www.exomol.com/}} opacity databases for simple ($<$10 atoms) molecules with spectral features near 3.17\,$\mu$m, leading us to consider HCN \citep{barber2014HCN}, NH$_3$ \citep{coles2019NH3}, C$_2$H$_2$ \citep{chubb2020C2H2}, C$_2$H$_4$ \citep{mant2018C2H4}, O$_3$ \citep{gordon2022hitran2020}, and C$_3$ \citep{lynasgray2024C3}; see the bottom panel of Fig.~\ref{fig-app:frspec} for the cross sections.
We consider these species in a free chemistry retrieval using PLATON, to determine if any of them are preferred by the data.\footnote{Note that most of these species are already included in PLATON's equilibrium retrievals, which considers e-, H, H-, He, C, N, O, Na, Fe, Ca, Ti, K, Ni, H$_2$, N$_2$, O$_2$, OH, CO, NO, SiO, TiO, VO, HCN, CH$_4$, CO$_2$, H$_2$O, H$_2$S, NH$_3$, PH$_3$, NO$_2$, SO$_2$, O$_3$, C$_2$H$_2$, and FeH \citep{Zhang2025PLATON}. However, here we are allowing our chosen species' abundances to vary freely, unlike in the equilibrium retrievals. }  
We also include H$_2$O \citep{polyansky2018H2O}, CH$_4$ \citep{yurchenko2024CH4}, CO$_2$ \citep{yurchenko2020CO2}, CO \citep{li2015CO}, SO$_2$ \citep{underwood2016SO2}, and H$_2$S \citep{azzam2016H2S} in the retrieval, as they are common molecules expected/detected in exoplanet atmospheres that have major absorption features in the 3-5 $\mu$m region. In addition, as with the equilibrium chemistry retrievals, we include the stellar radius, planet mass, planet radius (at 1 bar), an opaque pressure level, and an offset between NRS1 and NRS2 as free parameters, with the same priors as in Table \ref{tab:platonparams}, leading to 17 total parameters in the full retrieval (12 free chemical abundances).
For the chemical species abundances we consider a log-uniform prior with bounds of 10$^{-12}$ to 0.5, assuming a H$_2$/He background\footnote{For O$_3$ and C$_3$, which are more likely to form in an H-depleted atmosphere, we use H$_2$/He as a proxy for a low mean molecular weight background gas rather than an assumption of chemistry.}. 
To test the preference for any specific set of molecules, we also conduct retrievals without certain molecules, up to removing all but the common species, and compare their $\ln Z$ values.  
We use the \texttt{pymultinest} sampler and 1,000 live points and apply our retrieval framework to the combined \texttt{Tiberius} and \texttt{Eureka!} reductions only, since they have higher SNR than the individual visits.

\begin{figure*}
    \centering
    \includegraphics[width=0.98\linewidth]{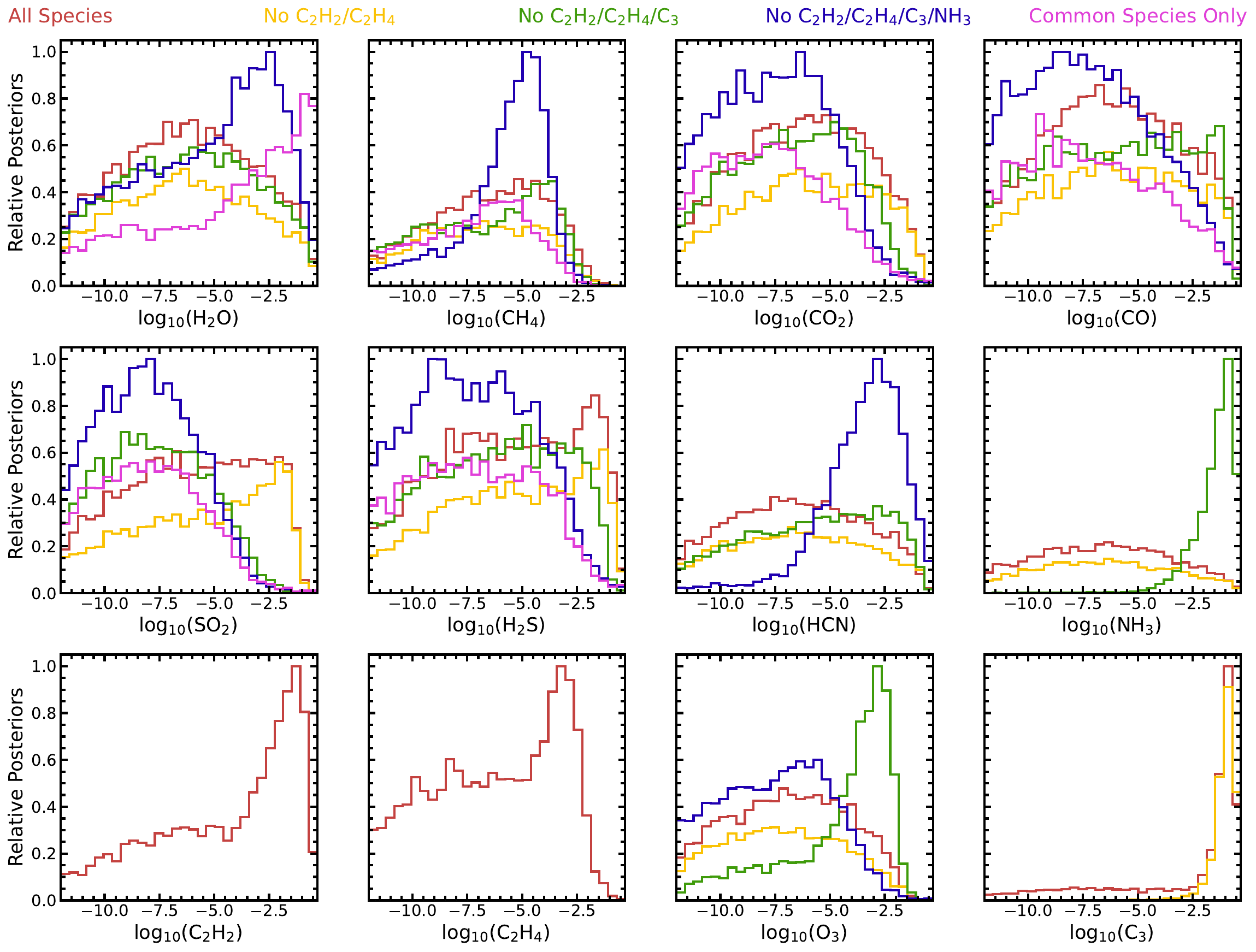}
    \caption{Retrieved abundance posteriors from our free retrieval analysis of the combined \texttt{Tiberius} reduction for our investigation of the 3.17\,$\mu$m feature. Various combinations of species were considered, with the same color-coding as in Figure \ref{fig-app:frspec}. }
    \label{fig-app:fram}
\end{figure*}

\begin{figure*}
    \centering
    \includegraphics[width=0.98\linewidth]{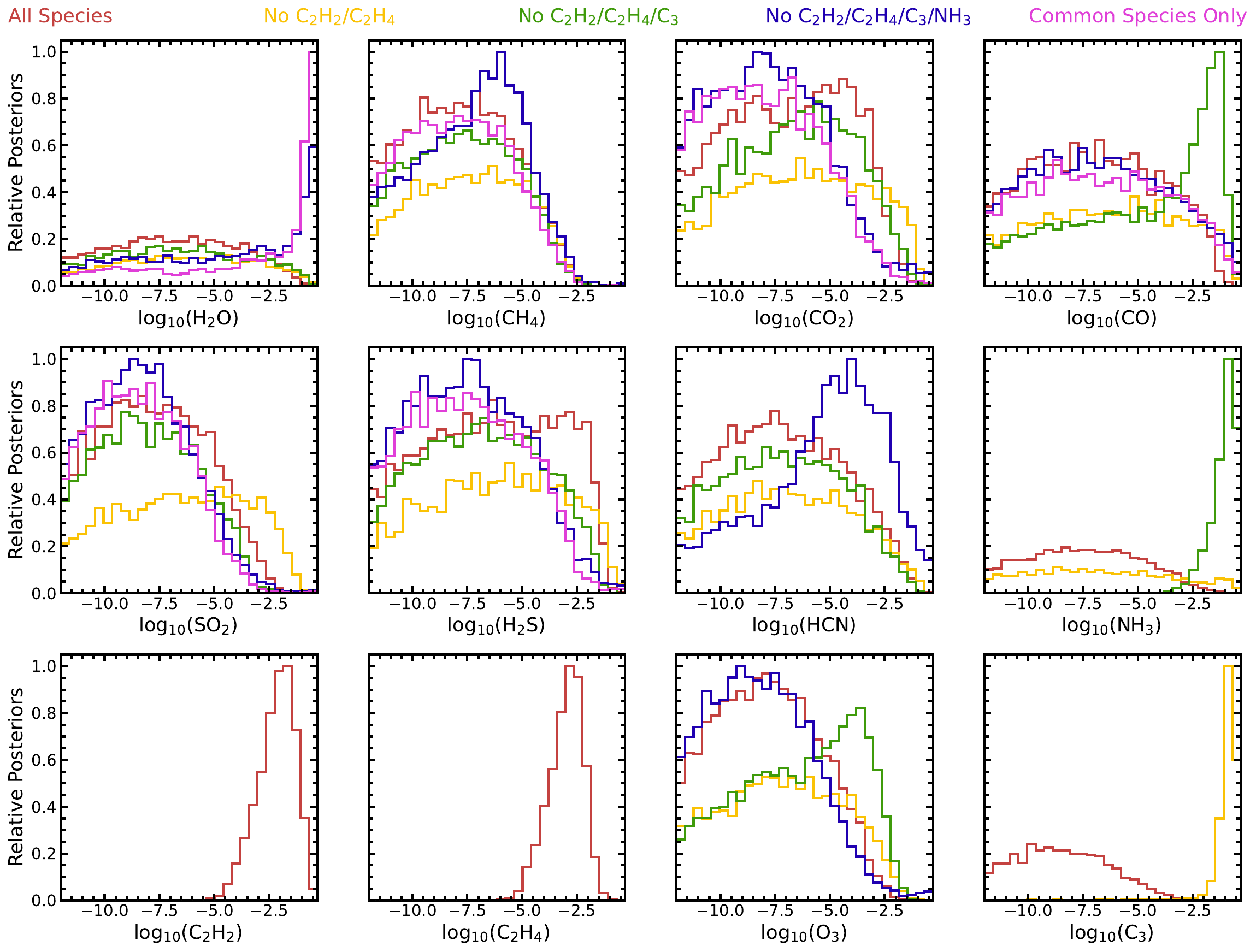}
    \caption{Same as Figure \ref{fig-app:fram} but for the \texttt{Eureka!}\ reduction.}
    \label{fig-app:frnw}
\end{figure*}

For our retrievals that include all 12 considered chemical species, we find that most of the abundances are completely unconstrained for both reductions, except for C$_2$H$_2$ and C$_2$H$_4$ (Fig.~\ref{fig-app:fram} and Fig.~\ref{fig-app:frnw}). 
For the \texttt{Tiberius} reduction, these two species are only slightly more constrained than the others, with lopsided posteriors peaking towards higher abundances but significant tails towards lower abundances. 
The posterior of C$_3$ is more constrained, with a narrow peak at the upper limit of the abundance prior, but still exhibits a tail towards lower abundances. Meanwhile, the retrieval on the \texttt{Eureka!} reduction more strongly constrains C$_2$H$_2$ and C$_2$H$_4$, but fails to constrain C$_3$. 
To investigate further, we conduct retrievals on the two reductions without C$_2$H$_2$ and C$_2$H$_4$ and compare their $\ln Z$ values with those of the full retrievals (Table \ref{tab:freechem}). 
While the $\Delta \ln Z$ is negligible for the \texttt{Tiberius} reduction ($<1$), for the \texttt{Eureka!} reduction the difference is significant ($>10$), indicating that the data strongly prefers the model with C$_2$H$_2$ and C$_2$H$_4$ to that without. With these two gases removed, C$_3$ is now tightly constrained for both reductions. We thus further remove C$_3$ and repeat the retrievals, resulting in a minor change in $\ln Z$ for both reductions ($\sim3$; Table \ref{tab:freechem}). These new results now show NH$_3$ to be more tightly constrained than the other molecules (Fig.~\ref{fig-app:fram} and Fig.~\ref{fig-app:frnw}), and therefore we further remove NH$_3$ and repeat the retrievals. For \texttt{Tiberius} this yielded a minor decrease in $\ln Z$ of $\sim2.7$, while for \texttt{Eureka!} $\Delta \ln Z \sim 7$, a more significant change (Table \ref{tab:freechem}). Finally, we conduct one more free retrieval with just the common species as a control, yielding negligible further decreases ($<2$) in $\ln Z$ for both reductions. 

Taken at face value, our free retrievals suggest that the 3.17\,$\mu$m feature--if it is caused purely by atmospheric gas absorption--can mainly be explained by the presence of one or more of C$_2$H$_2$, C$_2$H$_4$, C$_3$, and NH$_3$. However, this conclusion neglects additional astrophysical context about \planet\ and our retrieval results. First, the two reductions give different results for how much a given species contribute to the absorption. As the 3.17\,$\mu$m feature is more prominent in the \texttt{Eureka!} reduction than in the \texttt{Tiberius} reduction (Table \ref{tab:lnz}), $\Delta \ln Z$ values in the former are understandably larger (and therefore ``more significant'') than those in the latter. 
Second, the metallicity and mean molecular weight posteriors derived from the posteriors of the individual species in these free retrievals ([M/H]$\sim0-2.5$ and MMW$\sim2-5$\,g mol$^{-1}$, respectively) are much wider and lower than those from the equilibrium chemistry retrievals, which stems from the former attempting to fit the 3.17\,$\mu$m feature. 
Third, as discussed in recent works on the flaws of free retrievals and molecular ``detections'' \citep{welbanks2025arxiv}, our findings are limited by our list of considered species. Other species, including those without opacity data at these wavelengths but which may still have an absorption feature near 3\,$\mu$m, could also contribute. We therefore do not claim a detection of C$_2$H$_2$, C$_2$H$_4$, C$_3$, or NH$_3$, only that, out of our considered species, their absorption features best coincide with the 3.17\,$\mu$m feature in our spectra. 

It is uncertain whether some combination of C$_2$H$_2$, C$_2$H$_4$, C$_3$, or NH$_3$ can exist in the atmosphere of \planet at sufficient abundances to generate the 3.17\,$\mu$m feature. 
Near the equilibrium temperature of \planet, C$_2$H$_2$ and C$_2$H$_4$ can be produced via photolysis of CH$_4$ and subsequent recombination reactions, with higher metallicities and higher C/O leading to higher abundances \citep{moses2013,kawashima2019,mukherjee2025}.
However, in this scenario we would expect the 3.3\,$\mu$m CH$_4$ feature to be at least as prominent as the C$_2$H$_2$ and C$_2$H$_4$ features, which is not observed.
Alternatively, a He-dominated atmosphere that is H-poor and carbon rich (C/H$>$0.25) could thermodynamically prefer molecules like C$_2$H$_2$, C$_2$H$_4$, and C$_3$ so long as the atmosphere had little oxygen, as otherwise CO$_2$ would be the dominated C-bearing gas \citep{Cherubim2025}. 
NH$_3$ could plausibly be the dominant N species at the temperature of \planet\ depending on atmospheric mixing and internal heat \citep{fortney2020,mukherjee2025}. However, a large abundance would be required for it to be one of the only observed species in an otherwise featureless spectrum, suggesting highly enriched N compared to solar elemental ratios. Overall, additional observations of \planet\ are needed to confirm or reject the presence of these species and the validity and reproducibility of the 3.17\,$\mu$m feature.

\subsection{Likely atmospheric scenarios}

\begin{figure*}
    \centering
    \includegraphics[width=\textwidth]{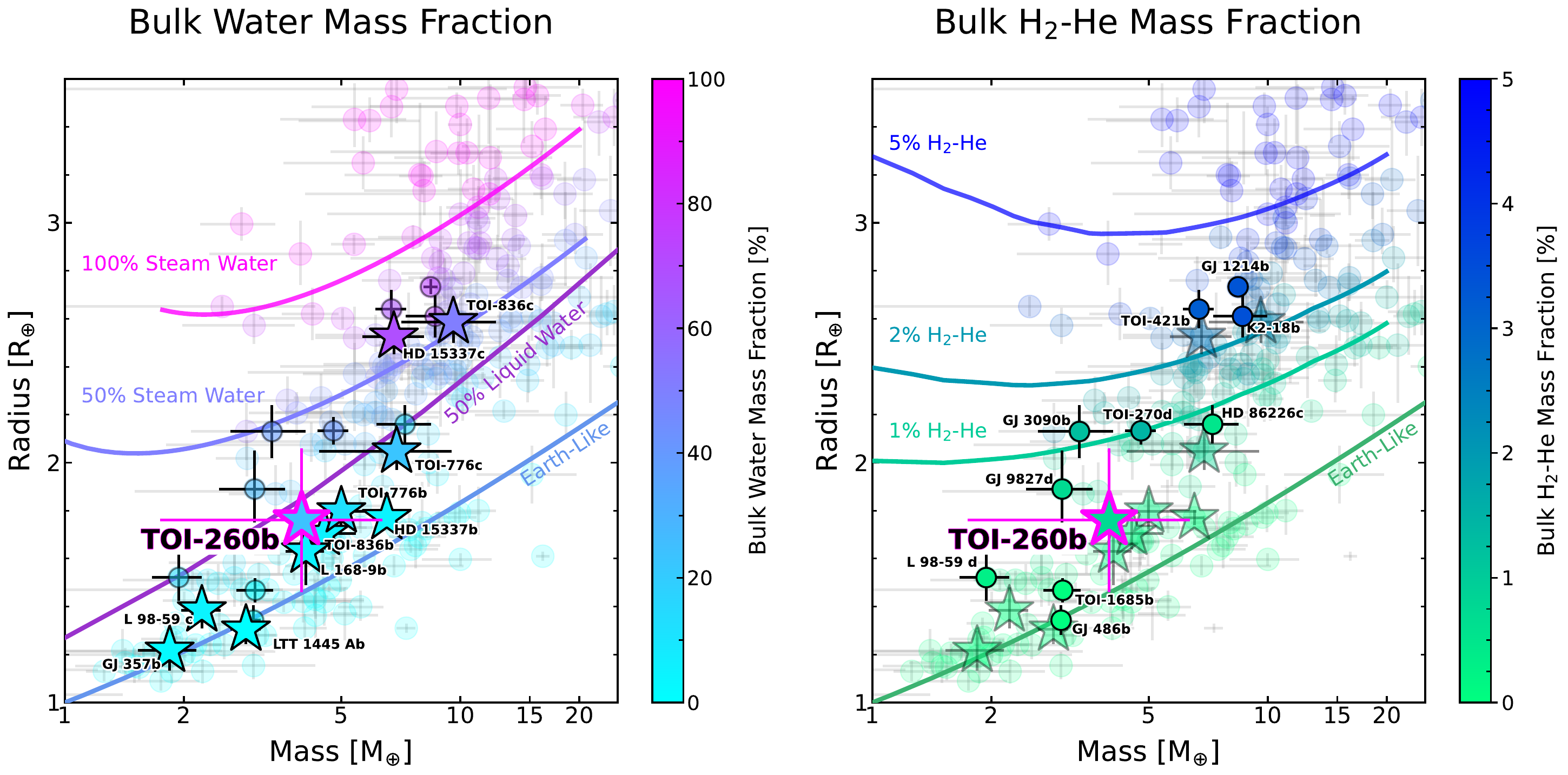}
    \caption{Small planet population with well-constrained mass and radii, sampled from the NASA Exoplanet Archive ($0.2<M_\mathrm{p}<30\,M_\oplus$).
    The data points are colored by (left) bulk water mass fraction and (right) bulk \ch{H2-He} mass fraction.
    The COMPASS targets are indicated by star symbols (labelled in the left panel), while a selection of other targets with published JWST/HST spectra are outlined (circles labelled in the right panel).
    We include a selection of theoretical models as solid lines: the Earth-like and 50\% liquid water curves are from \citet{zengMassRadiusRelationRocky2016}, the 1\%, 2\%, 5\% \ch{H2-He} curves are interpolated from the \citet{lopezUNDERSTANDINGMASSRADIUS2014} models, and the 50\%, 100\% steam water composition lines are interpolated from the \citet{aguichineMassRadiusRelationships2021} models, for 493\,K, age 5\,Gyr, and an Earth-like core.
    The color represents the bulk volatile content, calculated using the same method detailed in \citet{wallackJWSTCOMPASSNIRSpec2024}.
    The subject of this paper, \planet, is enlarged and highlighted in magenta.}
    \label{fig:mass-radius}
\end{figure*}

The observed mass and radius of \planet\ allow for a range of compositions (see Fig.~\ref{fig:mass-radius}), from rocky (Earth-like) to a nearly 50/50 water (steam)/rock world or a rocky core surrounded by a $\sim$1\% H$_2$/He envelope. Our largely featureless transmission spectra complement these constraints and show that the atmosphere needs to either be (i) high metallicity--similar to a water world--or (ii) low metallicity with abundant high-altitude aerosols. 
The atmosphere can also simultaneously host abundant aerosols and be metal-rich, or be completely absent, both of which would yield featureless spectra and are allowed by the observed mass and radius. On the other hand, if the 3.17\,$\mu$m feature is caused by atmospheric absorption, then a low molecular weight atmosphere is required. In that case, the volatile envelope would make up at most a few \% of the planet's mass (Fig.~\ref{fig:mass-radius}, right). 

\planet belongs to the relatively rare class of radius-valley ``dwellers'' that can be explained by a wide variety of internal structures and atmospheric compositions, and which also possess a diversity of spectra. 
While many radius-valley planets exhibit muted spectra \citep[e.g., TOI-776\,b and c, and GJ~3090\,b;][]{aldersonJWSTCOMPASSNIRSpec2025,TeskeJWSTCOMPASS2025,Ahrer2025}, others show features from a rich set of molecules \citep[e.g., TOI-270\,d]{Benneke2024arXiv} or appear to be steam worlds \citep[e.g., GJ~9827\,d;][]{piaulet-ghorayebJWSTNIRISSReveals2024}. \planet's spectrum is close to featureless aside from the 3.17\,$\mu$m bump, the origin of which is uncertain. As such, future investigations are needed to understand how \planet fits into the class of radius-valley planets.

\subsection{Future work}
The 3.17\,$\mu$m feature in the transmission spectrum of \planet is a mystery that can be investigated through further observations. 
Additional spectra taken at the relevant wavelengths, potentially with a different JWST/NIRSpec mode (e.g. G395M) or a different instrument altogether (e.g. NIRCam) may allow us to discriminate between instrument systematics and a real atmospheric signal. 
More generally, additional observations would be needed to ascertain the metallicity of the atmosphere and whether it could be He-rich, as was discussed in the previous section. 
This can be accomplished through e.g., looking for H and He outflows at Lyman-$\alpha$ and 1083 nm respectively, as was done recently for two other planets in the COMPASS sample to confirm significant H and He abundances in their upper atmospheres \citep[e.g. TOI-776\,c \& TOI-836\,c;][]{Loyd2025,Zhang2025He}. 
On the other hand, if the atmosphere is water-rich, as in the case of GJ 9827~d \citep{piaulet-ghorayebJWSTNIRISSReveals2024}, then follow-up JWST observations with NIRISS SOSS may be appropriate. 
Finally, all of the aforementioned current and future atmospheric observations would benefit from more precise mass and radius measurements, motivating observations of additional transits at high SNR and radial-velocity measurements.

\section{Summary \& Conclusions} 
\label{SECTION-7:conclusions}

We have presented a 3--5\,$\mu$m transmission spectrum of \planet, using two transit observations with JWST/NIRSpec G395H.
We summarize the key findings of this work below.
\begin{enumerate}
    \item We used two independent pipelines to reduce and fit the data, finding the transmission spectra to be consistent. The two visits afforded spectra which were similar in shape, albeit showing different detector offsets.
    \item Non-physical modeling confirmed that an offset between the two detectors was preferred over a flat line.
    We also observed evidence in favour of a Gaussian feature in NRS1 for some of the reductions and visits.
    We speculate as to the origin of this feature: while residual systematics are a probable cause, we also investigate possible atmospheric absorbing species.
    \item Using the PLATON retrieval framework, we rule out atmospheric metallicities below $\sim200\times$\,solar (assuming the opaque pressure level is $P_\mathrm{opaque}\gtrsim2.5$\,mbar), with the combined visits.
    \item The remaining atmospheric scenarios include a low metallicity atmosphere with a high aerosol layer, high-metallicity or no atmosphere. 
\end{enumerate}

With this work, we add to the sample of observed radius-valley sub-Neptunes, and provide atmospheric constraints towards refining our understanding of small-planet formation.
Further observations will be required to distinguish between the remaining atmospheric scenarios for \planet.

\section*{}
The data products associated with this work can be found at the following Zenodo repository: \href{10.5281/zenodo.18747984}{10.5281/zenodo.18747984}.
We thank the anonymous referee for their careful review of this work.
This work is based on observations made with the NASA/ESA/CSA James Webb Space Telescope. 
The data were obtained from the Mikulski Archive for Space Telescopes at the Space Telescope Science Institute, which is operated by the Association of Universities for Research in Astronomy, Inc., under NASA contract NAS 5-03127 for JWST. 
The specific observations analyzed can be accessed via \dataset[DOI: 10.17909/3had-nw24]{https://doi.org/10.17909/3had-nw24}.

These observations are associated with program \#2512. 
Support for program \#2512 was provided by NASA through a grant from the Space Telescope Science Institute, which is operated by the Association of Universities for Research in Astronomy, Inc., under NASA contract NAS 5-03127.
This work is funded in part by the Alfred P. Sloan Foundation under grant G202114194.
Support for this work was provided by NASA through grant 80NSSC19K0290 to J.T. and N.L.W. 
This material is based upon work supported by NASA’s Interdisciplinary Consortia for Astrobiology Research (NNH19ZDA001N-ICAR) under award number 80NSSC21K0597.
This work benefited from the 2022 and 2023 Exoplanet Summer Program in the Other Worlds Laboratory (OWL) at the University of California, Santa Cruz, a program funded by the Heising-Simons Foundation. 

Co-Author contributions are as follows: 
A.M. led the data reduction, analysis and write-up. 
P.G. led the modeling studies and write-up. 
N.L.W. contributed an additional independent reduction. 
M.L-M. and N.L.W contributed in the interpretation of results. 
D.O. and D.D. contributed data reduction and analysis of the ancillary CHEOPS and TESS data. 
All authors read and provided comments and discussion to improve the quality of the manuscript.

\software{
\texttt{batman} \citep{kreidbergBatmanBAsicTransit2015},
\texttt{emcee} \citep{DFM2013},
\texttt{Eureka!} \citep{bellEurekaEndEndPipeline2022},  
\texttt{ExoTiC-LD} \citep{grantExoTiCLDThirtySeconds2024}, 
\texttt{Matplotlib} \citep{Hunter2007},  
\texttt{NumPy} \citep{harris2020}, 
\texttt{PandExo} \citep{Batalha2017}, 
\texttt{PLATON} \citep{Zhang2019PLATON,Zhang2020PLATON},
\texttt{pyLDTK} \citep{Parviainen_2015_code},
\texttt{pymultinest} \citep{Buchner2014pymultinest}, 
\texttt{SciPy} \citep{Virtanen2020}, 
\texttt{Tiberius} \citep{kirkRayleighScatteringTransmission2017, kirkACCESSLRGBEASTSPrecise2021}.
}

\bibliography{01_coauthors,02_meech}{}
\bibliographystyle{aasjournal}

\appendix
\section{CHEOPS and TESS Data Analysis}
\label{APPENDIX-2:cheops}
TOI-260 was first observed by TESS in Sector 3 in September 2018 and then again in Sector 42 in September 2021. 
Following up on the initial detection of the planet candidate TOI-260.01 in the September 2018 TESS sector, we observed the system on September 12 and September 26 2021 with the European Space Agency's CHaracterizing ExOPlanets Satellite \citep[CHEOPS;][]{broegCHEOPSTransitPhotometry2013,benzCHEOPSMission2021} as part of its Second Announcement of Opportunity (AO-2) campaign. 
Unfortunately, both epochs missed the predicted transits because the ephemerides were not sufficiently accurate; a problem that we could only confirm after the TESS Sector 42 observations became available. 
With recomputed transit ephemerides using the TESS Sectors 3 and 42, we proposed for CHEOPS time again in its Third Announcement of Opportunity (AO-3), and obtained a successful third transit of TOI-260.01 on September 25 2022.  
Here we present the combined analysis of the three transit epochs (TESS Sector 3 and 42, and CHEOPS AO-3) and provide updated transit parameters and ephemerides.

The CHEOPS observations were processed by the CHEOPS Data Reduction Pipeline \citep[DRP;][]{hoyerExpectedPerformancesCharacterising2020}, which calibrates and corrects for instrumental and environmental effects, such as bias, gain, and flat-fielding. 
The DRP performs three main functions: calibration, correction, and photometry. 
The calibration phase consists of standard CCD reduction, including corrections for bias, gain, dark current, and flat-fielding. 
The correction phase accounts for smearing, bad pixels, depointing, pixel to sky mapping, and background and stray light. 
Finally, the DRP produces aperture photometry for radii from 15 up to 40 pixels.

\begin{figure}[t]
    \centering

    \includegraphics[width=0.8\linewidth]{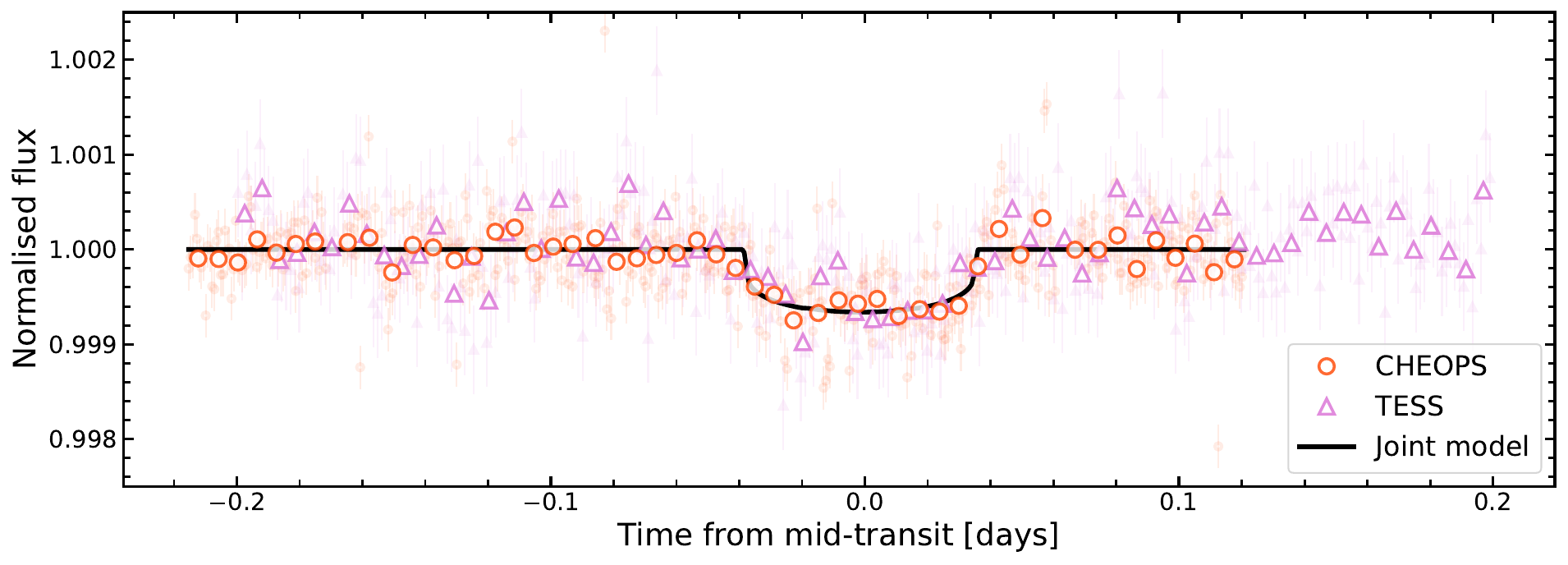}
    \caption{Phase-folded CHEOPS (red) and TESS (purple) and  light curves. The data binned to 10\,mins (open markers) are plotted for visual purposes only, while the unbinned data used for the fitting are plotted in the background.
    Overplotted in black is the best joint-fit \texttt{batman} transit light curve model.}
    \label{fig-app:CHEOPS_TESS_joint}
\end{figure}

\begin{table}
    \centering
    \caption{Retrieved astrophysical parameters from the best joint-fit of the CHEOPS and TESS light curves.}
    \begin{tabular}{l l}
    \toprule
        Parameter & Fitted value \\
        \midrule
         $T_\mathrm{mid}$\,[BJD] & $2459847.6884\pm0.0008$ \\
         Period\,[d] & $13.4759\pm0.00001$ \\
         $R_p\,[R_*]$ &  $0.0231\pm0.0005$ \\
         Transit duration [hrs] & $1.85\pm0.06$ \\
         \bottomrule
    \end{tabular}
    \label{table-app:cheops_params}
\end{table}

While the TESS light curves did not require detrending after using the Operations Center (SPOC) Pre-search Data Conditioning Simple Aperture Photometry (PDCSAP) extractions \citep{jenkinsTESSScienceProcessing2016}, the resulting CHEOPS light curves had visible systematics and outliers. 
We used \texttt{pycheops} \citep{maxtedPycheopsLightCurve2023} to trim outliers and detrend the CHEOPS light curve. 
We began by trimming outliers, which we defined as points $5\sigma$ away from the median value.
Given that CHEOPS rolls about its optical axis as it observes, the shape of the PSF changes throughout an observation. 
Therefore, we detrended against multiple parameters including $(X,Y)$ pixel position and telescope roll angle as first, second, and third order sine and cosine functions. 
We also detrended CHEOPS observations against background flux dominated by zodiacal light or scattered light from Earth, observation time and stellar contamination in the aperture. 
Finally, we checked nearby solar system objects in order to account for glint from these objects.
To avoid overfitting, we introduced each of these detrending parameters independently of one another and calculated the Bayes Factor with and without the parameter, as in \citet{trottaApplicationsBayesianModel2007}. 
This allowed us to determine whether each parameter was necessary for the model. 
We detrended for all of the above sources while simultaneously fitting a transit model with \texttt{pycheops} to avoid removing transit features.

We calculated limb-darkening coefficients as in \citet{oddoCharacterizationSetSmall2023}, \S5.4. 
We used the Python Limb Darkening Toolkit (PyLDTK; paper: \citealt{Parviainen_2015_paper}; code: \citealt{Parviainen_2015_code}) to calculate stellar limb-darkening coefficients for the power-2 limb-darkening law \citet{Hestroffer_1997,Maxted_2019} separately in both the TESS and CHEOPS bandpasses. 
We chose to keep limb-darkening coefficients in each bandpass constant through all fits and models for two reasons: to reduce uncertainties on other values, and because neither the TESS nor the CHEOPS photometry is sufficiently precise to allow a meaningful constraint on the limb-darkening coefficients.

We jointly fitted a \texttt{batman} transit model \citep{kreidbergBatmanBAsicTransit2015} to the detrended, phase-folded TESS and CHEOPS light curves, giving us tight constraints on the planet's mid-transit time, orbital period, transit duration, and transit depth. 
The fitting routine included finding an initial model with least-squares minimization, then initiating the model parameters in a full posterior exploration using the Markov Chain Monte Carlo sampler \texttt{emcee} \citep{DFM2013}. 
We initialized 40 walkers, ran the sampler for a burn-in phase of 500 steps, then ran for a total of 1500 steps. 
We checked convergence visually by inspecting the chains and corner plots.
The final transit model is shown phase-folded over both the CHEOPS and TESS data in Fig. \ref{fig-app:CHEOPS_TESS_joint}. 
The associated model parameter values are given in Table \ref{table-app:cheops_params}; only these parameters were directly fitted.

\section{Details of the Tiberius and Eureka! reductions}
\label{APPENDIX-1:reductions}

\begin{figure*}
    \centering
    \includegraphics[width=0.48\textwidth]{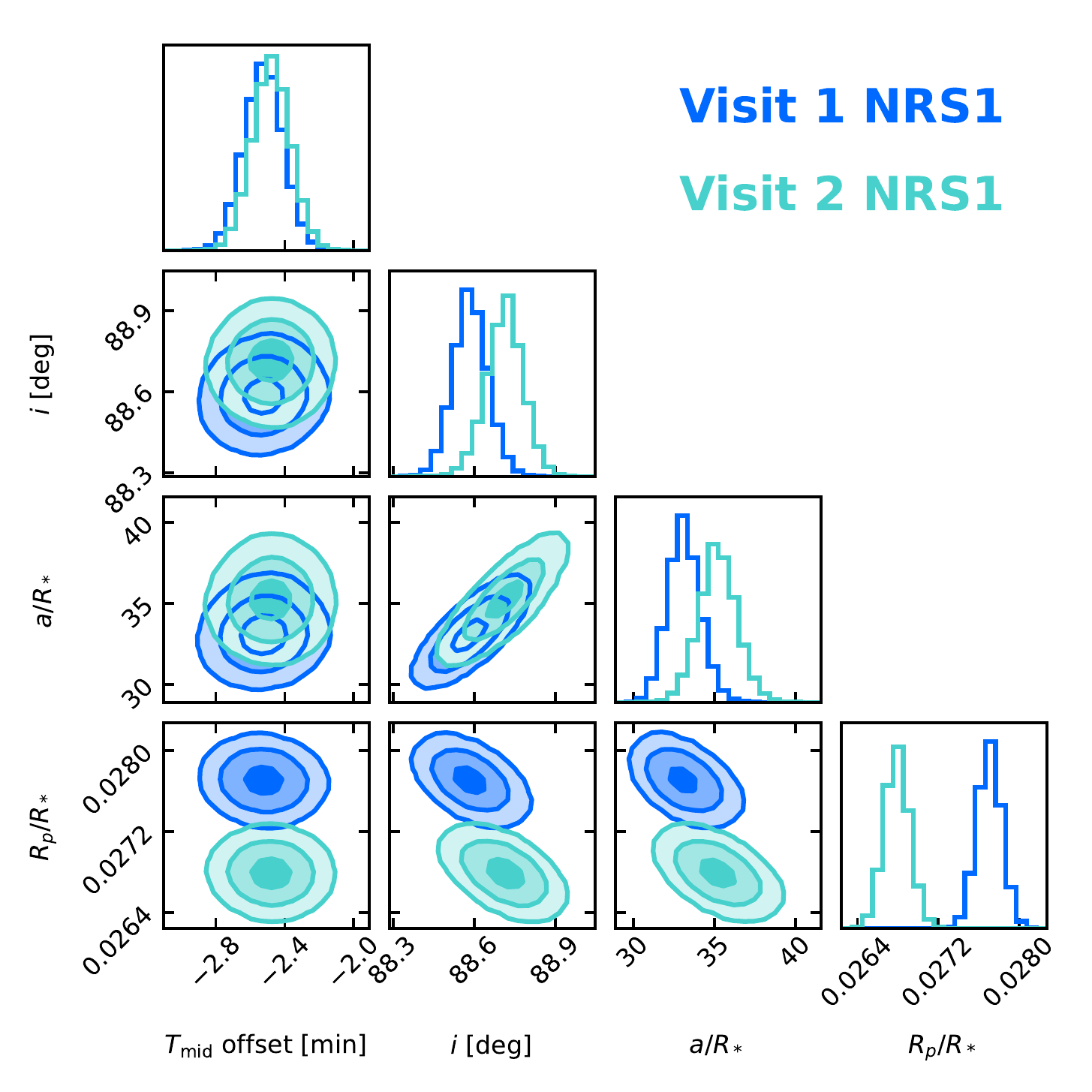}
    \includegraphics[width=0.48\textwidth]{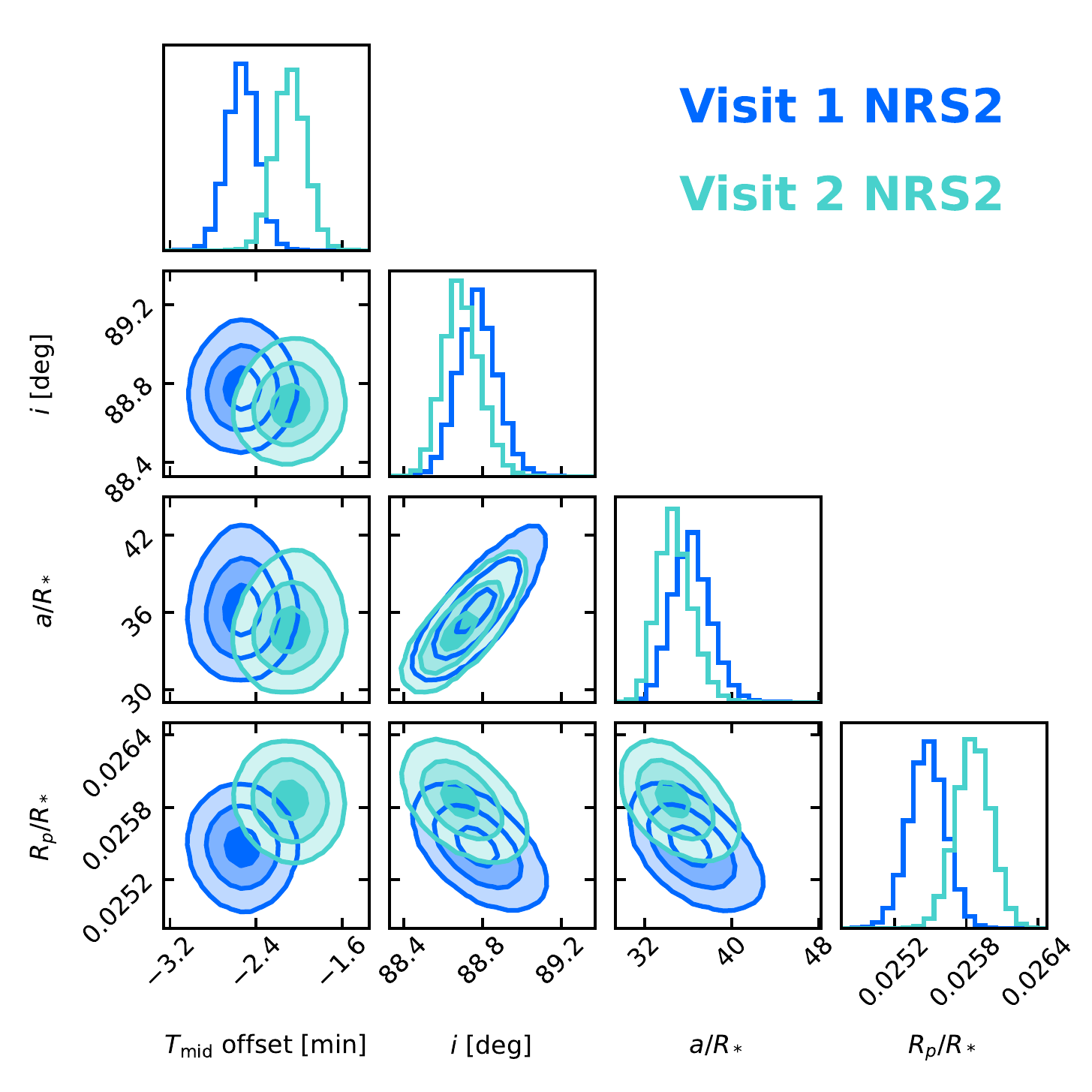}
    \caption{Posterior distributions (1, 2 and 3$\sigma$ confidence intervals) from the modelling of the individual \tiberius\ WLCs in \S\ref{SECTION-3:REDUCTION}. 
    The NRS1 results are shown on the left, and the NRS2 results on the right.
    The mid-transit time is given as the offset from the predicted mid-transit times provided in Table~\ref{table:retrieved-params}, in minutes.}
    \label{fig-app:wlc-corner}
\end{figure*}

\begin{figure}
    \centering
    \includegraphics[width=0.85\columnwidth]{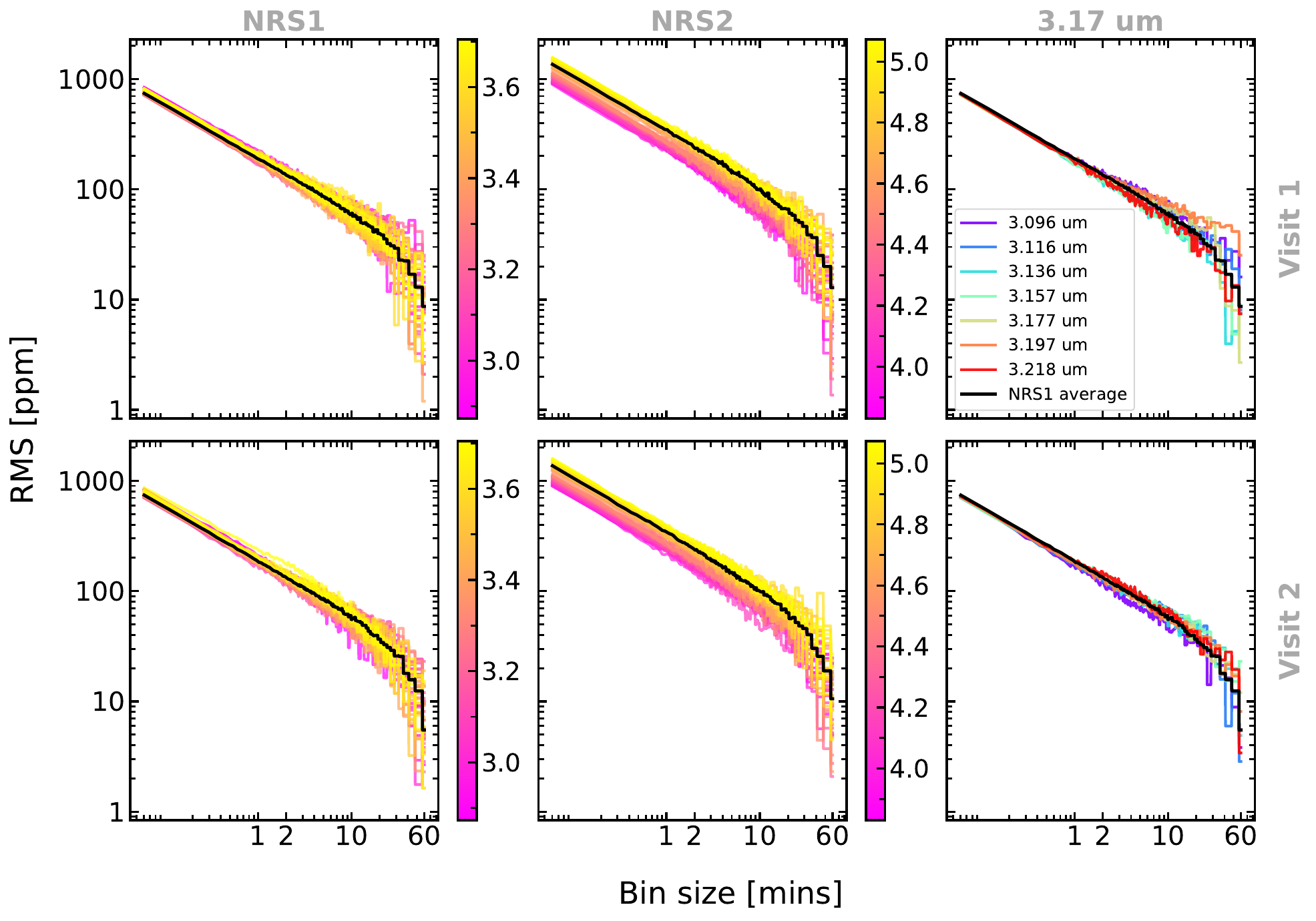}
    \caption{The RMS of the binned residuals from the \texttt{Tiberius} spectroscopic light curve fits, with the detectors NRS1 and NRS2 shown in the left and middle columns respectively. 
    The bin-center wavelengths (in microns) are indicated by the colorbar. Visit 1 and visit 2 are shown in the top and bottom rows respectively. The RMS of the binned residuals of the \texttt{Tiberius} spectroscopic bins around the $3.17\,\mu$m region are shown in the right column, wherein the average across \textit{all} spectroscopic bins in the NRS1 bandpass is shown in black for reference. }
    \label{fig-app:allan_SLC}
\end{figure}

In this Appendix, we provide further details of the independent reduction routines we used to process the JWST data, as presented in \S\ref{SECTION-3:REDUCTION}.
We summarise the final modelling choices for the individual light curve fits in Table~\ref{table-app:reduction-details}.
The posterior distributions from the \tiberius\ white light curve fits are shown in Fig.~\ref{fig-app:wlc-corner}.
As noted in \S\ref{SECTION-3:REDUCTION}, we trialed a number of WLC systematic models.
The recovered parameters from these tests are given in Table~\ref{table-app:WLC_params}.
The selected models are highlighted in bold in Table~\ref{table-app:WLC_params}; the astrophysical parameters are fixed to these values in the fitting of the SLCs. 
The RMS of the residuals of the SLC fits are shown in Fig.~\ref{fig-app:allan_SLC}.

{%

\begin{table}[]
    \centering
    \caption{Summary of the reductions described in \S\ref{SECTION-3:REDUCTION}. V1 and V2 are visit 1 and visit 2.}
    \label{table-app:reduction-details}
    \begin{tabular}{l c c}
    \toprule
         & \tiberius\ & \eureka\ \\
    \midrule
        
        \multicolumn{3}{c}{Stage 1 detector images reduction} \\
        \midrule
        \verb+jwst+ pipeline version & 1.13.4 & 1.11.4\\
        Group-level background subtraction? & Yes & Yes \\
        Outlier detection / sigma clipping & 4-sigma & 3-sigma\\
        
        \midrule
        \multicolumn{3}{c}{Spectral extraction} \\
        \midrule
        Standard/optimal extraction & Standard & Optimal\\
        Aperture full-width [pixels] & 8 & 10 for NRS1 V1, all other 8\\ 
        Spectral range extracted [microns] & NRS1: $2.748-3.722$; NRS2: $3.823-5.172$ & NRS1: $2.863-3.714$; NRS2: $3.812–5.082$\\
        \midrule
        \multicolumn{3}{c}{White light curves (WLCs)} \\
        \midrule
       
        Number of integrations trimmed & 992 (60\,mins) & 500 (30\,mins) \\
        WLC model & $S = p_1 + p_2
        \times T$ & $S= p_{1} + p_{2}\times T + p_{3}\times X + p_{4}\times Y$\\ 
        \midrule
        \multicolumn{3}{c}{Spectroscopic light curves (SLCs)} \\
        \midrule
        Wavelength bin width & 30 pixels &30 pixels \\
        Number of integrations trimmed & 992 (60\,mins) & 500 (30\,mins) \\
        SLC model & $S = p_1 + p_2
        \times T$ & $S= p_{1} + p_{2}\times T + p_{3}\times X + p_{4}\times Y$\\
        Average transit depth precision [ppm] & V1 25/ 45\,ppm; V2 25/ 45\,ppm$^\dagger$ & V1 24/ 40\,ppm; V2 23/ 41\,ppm\\
        
        \bottomrule
        
    \end{tabular}
\begin{tablenotes}
    \item $\dagger$ Values given for V1 (visit 1), V2 (visit 2) and NRS1/NRS2 respectively. 
\end{tablenotes}
\end{table}
}

\begin{sidewaystable*}
\renewcommand{\arraystretch}{1.2}
    \centering
    \begin{tabular}{l l c c c c c}
    \toprule
    Light curve  & Model & Trimming$^\dagger$ & $T_\mathrm{mid}$\,[BJD$_\mathrm{TDB}$] & $a/R_*$ & $i$\,[deg] & $R_\mathrm{p}/R_*$ \\
        
        \midrule
        \textbf{Visit 1 NRS1} & $\bm{S(t) = a_0 + a_1 t}$ & \textbf{60\,mins} & 
        $\bm{60264.939548{\substack{+0.000077\\-0.000079}}}$& $\bm{33.07{\substack{+ 1.04\\-0.96}}}$& $\bm{88.58\pm0.06}$  & $\bm{0.027706\pm 0.000120}$\\
        & $S(t) = a_0 + a_1 t + a_2 \exp{t}$ & None & $60264.939495{\substack{+0.000081\\-0.000082}}$ & $33.86{\substack{+1.17\\-1.07}} $&$88.63\pm0.07$ & $0.026915{\substack{+0.000147\\- 0.000149}}$\\
        & $S(t) = a_0 + a_1 t + a_2 X + a_3 Y + a_4 F$ & 60\,mins & $60264.939546\pm0.000078$& $32.98{\substack{+1.00\\-0.96}}$ & $88.58\pm0.06$ & $0.027723{\substack{+0.000121\\- 0.000120}}$\\

        \textbf{Visit 1 NRS2} & $\bm{S(t) = a_0 + a_1\times t}$ & \textbf{60\,mins} & 
        $\bm{60264.939544\pm0.000096}$& $\bm{36.21{\substack{+1.88\\-1.72}}}$ & $\bm{88.77\pm0.10}$ & $\bm{0.025472{\substack{+0.000153\\-0.000154}}}$\\
        & $S(t) = a_0 + a_1 t + a_2 t^2$ & None & $60264.939584{\substack{+ 0.000095\\- 0.000096}}$& $35.89{\substack{+ 1.85\\- 1.69}}$ & $88.76\pm0.10 $ & $0.025751{\substack{+ 0.000177\\- 0.000178}}$\\

        \textbf{Visit 2 NRS1} & $\bm{S(t) = a_0 + a_1 t}$ & \textbf{60\,mins} & $\bm{60278.415478{\substack{+0.000076\\-0.000077}}}$& $\bm{35.21\pm1.16}$ & $\bm{88.71\pm0.07}$ & $\bm{0.026788{\substack{+ 0.000124\\- 0.000122}}}$\\
        & $S(t) = a_0 + a_1 t + a_2 \exp{t}$ & None & $60278.415502\pm0.000076$& $35.68{\substack{+1.22\\- 1.18}}$ & $88.74\pm0.07$ & $0.026356 \pm0.000185 $\\

        \textbf{Visit 2 NRS2} &  $\bm{S(t) = a_0 + a_1\times t}$ & \textbf{60\,mins} & $\bm{60278.415752{\substack{+ 0.000099\\- 0.000100}}}$& $\bm{34.68{\substack{+1.69\\- 1.47}}}$ & $\bm{88.69{\substack{+ 0.10\\- 0.09}}}$ & $\bm{0.025859{\substack{+ 0.000146\\- 0.000148}}}$\\

        & $S(t) = a_0 + a_1 t + a_2 t^2$ & None & $60278.415659\pm0.000101$ & $34.58{\substack{+ 1.79\\- 1.50}} $ & $88.68{\substack{+0.10\\- 0.09}}$ & $0.025703{\substack{+ 0.000178\\- 0.000183}}$\\

        Joint NRS1 & $S(t) = a_0 + a_1 t$ & 60\,mins & $60264.939572\pm0.000059$ & $34.00\pm0.60$ & $88.64\pm0.04$ & $0.027259\pm0.000066$\\
        & & & $60278.415456{\substack{+0.000060\\-0.000059}}$\\
        
        Joint NRS2 & $S(t) = a_0 + a_1 t$ & 60\,mins & $60264.939545\pm0.000080$ & $35.25{\substack{+ 0.96\\- 0.89}}$ & $88.72{\substack{+0.06\\-0.05}}$ & $0.025680{\substack{+ 0.000083\\- 0.000086}}$\\
        & & & $60278.415768\pm0.000079$ \\
        
    \bottomrule
    \end{tabular}
    \caption{Fitted parameters from various trialed white light curve systematics models (see \S\ref{SECTION-3:REDUCTION}). The adopted parameters are given in the main body of the paper, Table~\ref{table:retrieved-params}, and re-printed in bold here for reference.}
    \label{table-app:WLC_params}
\begin{tablenotes}
    \item $\dagger$ Trimming refers to the exclusion of integrations at the \textit{beginning} of the observation.
\end{tablenotes}
\end{sidewaystable*}

From fitting the detectors and visits individually (\S\ref{SECTION-3:REDUCTION}), the recovered system parameters were somewhat discrepant.
In particular, the range of measured $R_\mathrm{p}/R_*$ was an order of magnitude larger than the 1-sigma uncertainties (Table~\ref{table:retrieved-params}).
We performed joint fits of the \tiberius\ light curves as an independent check, fitting the NRS1 light curves from both visits simultaneously, with shared $R_\mathrm{p}/R_*$, $i$, $a/R_*$, and distinct systematics models.
We used the same MCMC set-up and the same pre-transit trimming (60\,mins) described above.
Having zero-centered the time arrays on the predicted mid-transit time, propagated from the literature value (Table~\ref{table:retrieved-params}), we fit for distinct offsets $\Delta T_0$ for each visit.
The recovered parameters from the joint WLC fits are given in Table~\ref{table-app:WLC_params}, wherein $\Delta T_0$ has been propagated back to $T_\mathrm{mid}$ for ease of comparison.
For NRS1, the jointly fit $T_\mathrm{mid}$, $a/R_*$ and $i$ were consistent within 1-sigma with the individual fit values (Table~\ref{table:retrieved-params}), with a small improvement on the uncertainty, and were almost equal to the mean of the individual visit values.
For NRS2, the jointly fit astrophysical parameters were similarly in agreement with the individual fit values.

\begin{figure}
    \centering
    \includegraphics[width=0.96\linewidth]{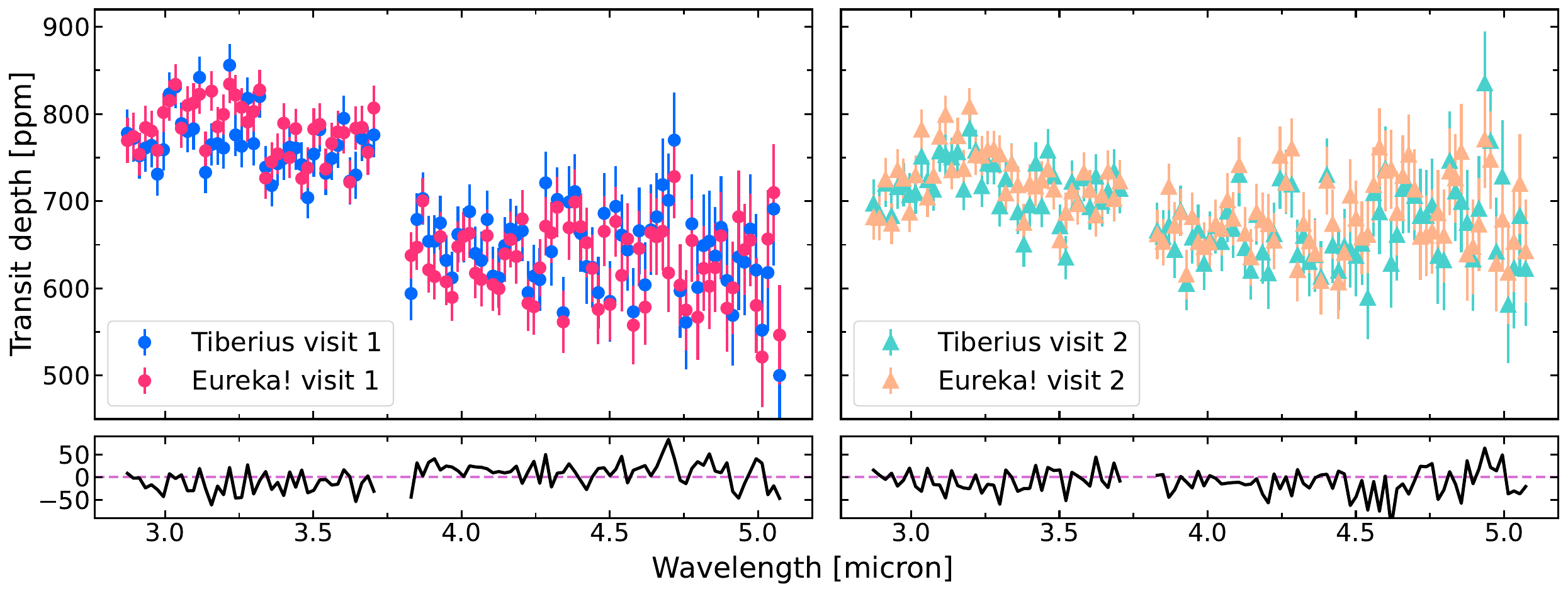}
    \caption{
    The transmission spectra of TOI-260\,b, extracted with \texttt{Tiberius} and \texttt{Eureka!} for visit 1 (left) and visit 2 (right).
    The difference between the reductions in ppm is shown below in black.  The spectra displayed here are identical to those in Fig.~\ref{fig:transmission_spectra}, where instead the reductions are shown in separate panels.}
    \label{fig-app:transmission_spectra_visits}
\end{figure}

We then fixed the astrophysical parameters to the jointly fit WLC values (Table~\ref{table-app:WLC_params}) and jointly fit the SLCs from each visit, again keeping the detectors separated.
The limb-darkening coefficients fixed to the values determined in \S\ref{section-3:redution_tiberius}.
We implemented distinct systematics models, such that there were five free parameters per joint spectroscopic light curve fit, the four linear coefficients and $R_\mathrm{p}/R_*$.
The resulting transmission spectrum was nearly identical to the weighted average transmission spectrum presented in \S\ref{SECTION-3:REDUCTION}, with an average difference of 1\,ppm and a maximum difference of 4\,ppm.
We are therefore confident that the discrepancies between the fitted system parameters have little bearing on the shape of the transmission spectrum.

To enable easier comparison between the separate reductions used for the final transmission spectrum in Fig.~\ref{fig:transmission_spectra}, we plot the reductions on the same panels in Fig.~\ref{fig-app:transmission_spectra_visits}.
We see excellent agreement between the reductions, with an average absolute difference in transit depth of 23\,ppm and 22\,ppm in visit 1 and 2 respectively, smaller than the average precisions.

\end{document}